\documentclass{jfm}
\usepackage{graphicx}
\usepackage{epstopdf, epsfig}
\usepackage{subcaption}
\usepackage{amsmath}

\shorttitle{Modelling a Hydrodynamic Instability}
\shortauthor{Zs. Varga, J.L. Hofmann and J.W. Swan}
\title{Modelling a Hydrodynamic Instability in Freely Settling Colloidal Gels}

\author{Zsigmond Varga,
  Jennifer L. Hofmann
 \and James W. Swan\corresp{\email{jswan@mit.edu}}}

\affiliation{Department of Chemical Engineering, Massachusetts Institute of Technology, Cambridge, MA 02139, USA}

\begin{document}

\maketitle

\begin{abstract}
Attractive colloidal dispersions, suspensions of fine particles which aggregate and frequently form a space spanning elastic gel are ubiquitous materials in society with a wide range of applications. The colloidal networks in these materials can exist in a mode of free settling when the network weight exceeds its compressive yield stress.  An equivalent state occurs when the network is held fixed in place and used as a filter through which the suspending fluid is pumped.  In either scenario, hydrodynamic instabilities leading to loss of network integrity occur.  Experimental observations have shown that the loss of integrity is associated with the formation of eroded channels, so-called streamers, through which the fluid flows rapidly. However, the dynamics of growth and subsequent mechanism of collapse remain poorly understood. Here, a phenomenological model is presented that describes dynamically the radial growth of a streamer due to erosion of the network by rapid fluid back flow.  The model exhibits a finite-time blowup -- the onset of catastrophic failure in the gel -- due to activated breaking of the inter-colloid bonds. Brownian dynamics simulations of hydrodynamically interacting and settling colloids in dilute gels are employed to examine the initiation and propagation of this instability, which is in good agreement with the theory. The model dynamics are also shown to accurately replicate measurements of streamer growth in two different experimental systems.  The predictive capabilities and future improvements of the model are discussed and a stability-state diagram is presented providing insight into engineering strategies for avoiding settling instabilities in networks meant to have long shelf lives.
\end{abstract}

\section{Introduction}
Colloidal gels are among the most abundant soft matter found in society. They are the components of everyday products including foodstuffs\citep{mezzenga2005understanding}, consumer care products\citep{hu2012novel}, cosmetics,  paints\citep{russel1989colloidal}, pesticides and proppants in oil and gas exploration\citep{bai2007preformed}. Additionally, careful control and intelligent design of particle gels is critical for several emerging materials applications, found in 3D printing inks, micro-fuel cells\citep{Gaponik2011} and membranes\citep{yang2004molecularly}. One of the most attractive engineering features of these space-spanning networks of attractive particles is their yield stress. Typically this is high enough to bear the material's own weight, but low enough to give flowability during use\citep{poon2002physics, zaccarelli2007colloidal}. One key design requirement is that the particles must not sediment appreciably during the product's 'shelf life', which might range from weeks to years. Despite this restriction, a majority of colloidal gels, which contain non-density matched particles, exhibit various, undesired settling behaviours such as streamer formation and network collapse. In a range of other industrial scenarios, such as in conformance control during hydrofracking or industrial filtration, an equivalent state to mechanical compression occurs when the suspending fluid is pumped through the colloidal gel fixed in place and network collapse leads to unrestricted fluid flow \citep{northcott2005dewatering, macminn2016large}. In either scenario, failure of the gel is equivalent to loss of utility of the product for these applications. Consequently, the ability to engineer and increase the stability of these elastic networks remains an important and prevalent issue for many industries. 

The theory of sedimentation of attractive colloidal dispersions was developed by \citet{buscall1987consolidation} describing the interplay of three forces that control the extent of sedimentation: the gravitational driving force, the viscous drag force associated with flow of liquid around and through the solid and the elastic stress developed in the network of particles. Sedimentation occurs when the gravitational force exceeds the local yields stress of the network resulting in three distinct zones of behaviour: the supernatant, the falling zone, and the consolidating zone. The supernatant is the particle-free region above the network that is formed following the onset of sedimentation, whereas the consolidating zone at the bottom is the region throughout which the local yield stress exceeds the network pressure above it. In the falling zone, the gravitational driving force is balanced only by the viscous drag due to local fluid back flow and all particles settle freely at a rate that, in theory, can be directly related to the dispersion's height profile. For a "tall" macroscopic sample, the majority of the particle network constitutes the falling zone and particles settle freely over experimentally relevant timescales. 

Scientific studies of gravitational collapse of gels have in the last three decades focused on examining the settling behaviour of model colloidal dispersions. New Insights into colloidal aggregation and rearrangements under the influence of gravity could ultimately provide a thorough understanding of real aggregating systems\citep{huh2007microscopic}, improve pressure-filtration driven fine solids separation processes\citep{buscall1987consolidation}, and elucidate the effects of gravity on the kinetics of arrested phase separation\citep{bailey2007spinodal, kim2013gel} and on diseases related to protein aggregation: sickle cell anaemia, Alzheimer's disease and amyloid fibril growth \citep{growthclark1987phase}.

Just as seen in industrial applications, the long-time structural integrity of an experimental gel, if not exactly density-matched, is constrained by the gravitational stress exerted by its own weight and the network may undergo gravitational collapse\citep{starrs2002collapse, bailey2007spinodal, kamp2009universal}. The most common and ultimate metric characterizing the macroscopic feature of the collapse process is the time evolution of the height profile, $h(t)$ of a gel. Measurements of $h(t)$ are used to determine the characteristic timescale of the process, $t_d$, and to quantify observed power-law or exponential decays of height profiles\citep{weeks2000formation, teece2014gels, harich2016gravitational}. The collapse dynamics of colloidal dispersions with long-ranged attractions, relative to the primary particle radius, are relatively well understood. Here, the network is transient, continuously coarsens and sediments over time exhibiting dynamics of a phase separation process analogous to spinodal decomposition\citep{teece2011ageing}. In contrast, in the case of gels with short-ranged interaction, extensive experimental investigations performed in the past decades have shown that after a seemingly arbitrary quiescent period, the dynamics of settling and compaction of the gel may proceed by widely different means depending on the range and strength of the particle interactions, and on the particle concentration within the gel\citep{secchi2014time}.  

"Strong" gels exhibit slow, uniform compression that haults once the yield stress of the now more compact network exceeds its own weight\citep{manley2005gravitational, brambilla2011highly}.   In contrast, a "weak" gel initially shows a similar slow, uniform compression for a duration equal to the delay time $t_d$, before suddenly undergoing a rapid and catastrophic collapse\citep{allain1995aggregation, poon1999delayed, starrs2002collapse, kilfoil2003dynamics, blijdenstein2004scaling, huh2007microscopic, kamp2009universal, bartlett2012sudden, harich2016gravitational}. This has been observed in a wide variety of systems with short-ranged attraction, and the response appears to be a universal feature of "weakness". The distinction between strong and weak gels is purely based on the temporal dependence of the position of the meniscus separating the freely settling gel from the supernatant phase - a macroscopic observable, carrying very limited amount of information\citep{teece2014gels}. Microscopic insights of how the network evolves in time, how it transmits stress and what distinguishes strong from weak gels on the microstructural level are still elusive.  Understanding how settling gels can be turned from weak into strong would facilitate the design of colloidal gel products with longer shelf-lives and the ability to prevent delayed collapse within the desired user time frame.

Experimental colloidal gels already contain on the order of $10^6$ particles, and materials in actual industrial applications can contain several orders of magnitude more. Treating such large systems in a theoretical fashion therefore is, at present, only possible with approximate approaches, which describe the height profile via one dimensional transport equations\citep{macminn2016large}. Nonetheless, in the case of strong gels, a continuum model for the collapse rate has been developed based on the theory of poro-elasticity. It takes into account the resistance to compression arising from a combination of the fluid pressure and the elasticity of the network and successfully captures the full collapse behaviour\citep{manley2005gravitational}. In contrast, for delayed collapse to date no widely accepted theoretical framework has emerged to account for the process\citep{kilfoil2003dynamics, huh2007microscopic, harich2016gravitational}. Among other complicating factors and shortcomings, the poro-elastic theory fails due to the highly non-linear nature of the rapid collapse \citep{manley2005gravitational}. Alas, without a firm understanding of the collapse dynamics, the actual stability of many products remains unpredictable and uncertain.

While in general experiments cannot observe the microstructure of the catastrophic collapse of these networks, a few careful studies have provided insight on the relationship between changes in microstructure and macroscopic collapse that can drive further theoretical development. Experiments conducted using dark-field microscopy imaging have reported landmark observations of the dramatic hydrodynamic instabilities that precede the sudden and rapid collapse\citep{poon1999delayed, starrs2002collapse}. The experiments observe the nucleation and growth of a large channel that is absent of particles, which provides a path for significant fluid back flow through it, a so-called "streamer". It grows in radius, and eventually also spans the height of the gel column, causing a small "eruption" at the interface between the supernatant and the settling gel\citep{senis2001settling}, when catastrophic loss of network integrity occurs. It has been hypothesized that compacting gel fragments breaking off from the top interface fall through the dilute network, and are responsible for the creation of the streamers that generate the hydrodynamic back flow and subsequent instability\citep{harich2016gravitational}. However, recent ghost particle velocity studies of collapsing gels\citep{secchi2014time} that were able to measure the hydrodynamic velocity pattern before rapid collapse, have however reported the onset of two vertical streamers originating in the bulk of the sample. They progressively expand, providing a path for the back flow of the fluid. Soon after, the breakdown of the gel becomes extremely fast, rapidly leading to the full disruption of the gel structure. It is thus evident, that fluid flow, the drag and hydrodynamic interactions exerted by the solvent on the particle network is crucial to understanding the collapse of freely settling gels. 

Careful confocal microscopy studies of the association and dissociation processes of individual particles and gel strands in the network structure have also shown that there is a direct link between the macroscopic mean delay time $t_d$ and the lifetime of an individual colloid-colloid bond\citep{gopalakrishnan2006linking, teece2014gels}. As a result simple phenomenological models have been developed that describe the collapse process in terms of a number of sticky inter-particle bonds undergoing sequential activated bond breaking and leading to a loss of network integrity. These models can relate the microscopic dynamics to the macroscopic settling with some success, albeit they have no ability to predict the onset of delayed collapse\citep{kamp2009universal, teece2014gels}. Additionally, no consideration has been given to the role of hydrodynamics in driving the collapse process so that a theory explaining the observed dynamics in terms of controllable network parameters remains to be developed. 

Given the experimental limitations and lack of comprehensive theories, dynamic computer simulations provide an invaluable tool to study the microscale dynamics of the hydrodynamic instability preceding collapse. A computer model is able to offer detailed particle level information of the entire gel network throughout the settling process, provided the simulation is able to capture the correct physical processes involved during collapse. Recently we have shown that the collective dynamics facilitated by the presence of long-ranged hydrodynamic interactions enable gelation and play a critical role in setting their relaxation dynamics\citep{varga2015hydrodynamics, varga2016hydrodynamic, varga2017normal}. Computer models that only include solvent-mediated hydrodynamic effects at the one-body level, i.e. Stokes drag, fail to reproduce the experimentally observed gel mechanics. This will especially be the case for the collapse of freely settling gels, which appears to be tripped by a hydrodynamic instability. For instance, in one example of the limited number of computer studies on settling gels, experimentally observed dense clusters that form during gel collapse have not been reproduced in simple Brownian dynamics simulations lacking hydrodynamic interactions\citep{harich2016gravitational}. In another case, where gels were confined vertically in capillary tubes to within one gravitational length in experiments, Brownian dynamics simulations neglecting hydrodynamic interactions qualitatively reproduced the sedimentation profile. However, comparison with observations of bulk systems and the process of delayed collapse suggests that there is a fundamental difference in mechanism between these confined systems and bulk measurements of industrial relevance, which exhibit settling rates several orders of magnitude slower\citep{razali2017effects}.

Here we present a comprehensive study of the hydrodynamic instability leading to collapse of freely settling colloidal gels, combining a new theory with computer simulation studies and comparison to experiments. We propose a microstructural model for the hydrodynamic instability comprised of nucleation and growth of streamers driven by network erosion from fluid back flow that leads to rapid gravitational collapse. The model relates the delay period prior to streamer blowup and gel collapse to various properties of the gel network, including particle volume fraction, the attraction range, interaction strength between particles relative to thermal forces, and gravitational strength. We study freely settling colloidal gels using Brownian dynamics simulations of attractive, hydrodynamically interacting particles in order to examine the formation and growth of these hydrodynamic instabilities. The settling velocity and streamer volume are measured as they evolve over time, and a critical point for the onset of rapid growth in both quantities is identified. The settling process is well described by our theory. The model is also compared with two different experimental systems and is found to accurately predict the collapse dynamics.

This article is organized as follows. In $\S$\ref{sec:model} we present the details of the   microstructural model for network erosion and streamer growth in a model colloidal gel and arrive at a scaling law for the blowup time, i.e. the onset of catastrophic collapse as a function of network parameters. Next, in $\S$\ref{sec:validation} we present extensive results to validate our theory by comparing the model predictions to the collapse dynamics of freely settling gels in simulations and published experiments. Section $\S$\ref{sec:discussion} discusses a new conceptual framework for how to think about network stability and how the model can aid the engineering of these materials. We highlight potential areas of improvement and subsequently conclude our work in $\S$\ref{sec:conclusions}.

\section{A model for network erosion and streamer growth leading to gravitational collapse}
\label{sec:model}
The model considers a system spanning percolated elastic gel network of attractive, spherical colloidal particles of radius $a$. The gel has a volume fraction $\phi$ and the tortuousness and porosity of this kinetically arrested material is characterized by the fractal dimension $d_f$. The short-ranged attractive well has depth $U$ and width $\Delta$. The particles have a thermal energy $kT$, density mismatch $\Delta\rho$ with the suspending fluid of viscosity $\eta$, and settle freely under the effect of a uniform gravitational acceleration $g$.  This physical scenario is characterized concisely by only 5 dimensionless quantities, summarized in table \ref{tab:groups}.

\begin{figure}
  \centerline{\includegraphics[width=\columnwidth]{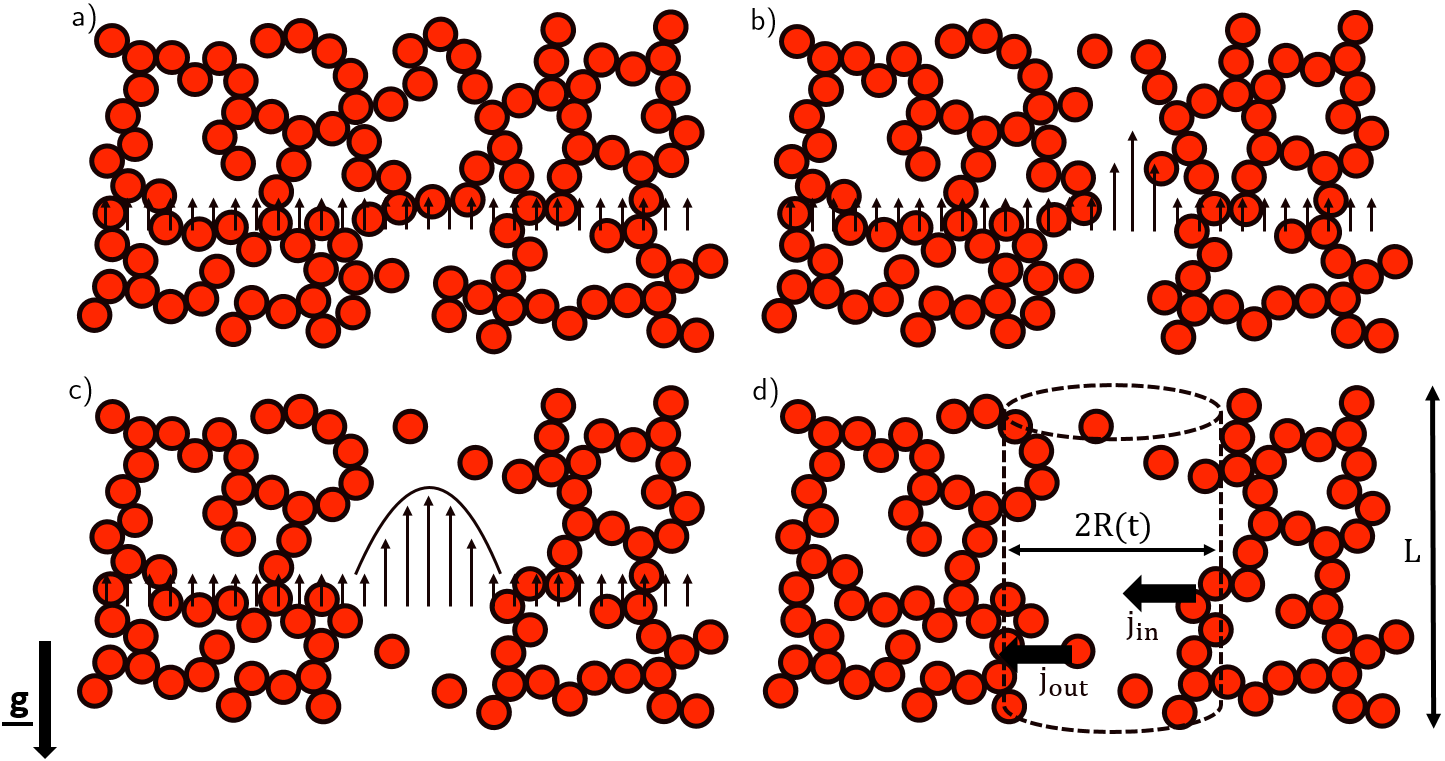}}
  \caption{The micromechanical model considers a particle gel network freely settling in a gravitational field \textbf{g}. a) As the particles move downward, to conserve mass, fluid will flow upward through the pores(black arrows). b) Due to local density fluctuations the fluid will find a path of least resistance, where increased back flow will nucleate a streamer that spans the network. c) As the back flow increases, the fluid exerts drag on the particles at the interface leading to an activated rate of erosion and growth of the streamer. d) The streamer is modelled as a cylindrical channel with evolving radius $R$ in a gel of size $L$. The growth rate will depend on the relative magnitude of particle fluxes away from ($\j_\mathrm{in}$) and into the network ($j_\mathrm{out}$).\label{fig:pore_model}}
\end{figure}

In the free falling zone during sedimentation the network does not experience any effects from interactions with the supernatant interface and is unaware of the build up of the dense cake forming in the consolidating zone\citep{buscall1987consolidation}. As the particles move downward, to conserve mass, fluid will flow upward through the pores of the gel. Initially, under the imposed constant hydrostatic pressure gradient, $\left|g\Delta\rho\right|=\left|\nabla p\right|$, fluid flow will be uniform. Due to local density fluctuations and further restructuring processes, local differences in the permeability of the network will be set up. This results in a burst in local fluid velocity relative to the mean and the fluid will nucleate a path of least resistance. The increased volumetric flow rate of back flow through this initial channel exerts hydrodynamic drag on the particles in the network and leads to activated breaking of particle bonds, locally eroding the gel. This erosion leads to the widening of a largely cylindrical streamer and greater fluid back flow locally that further accelerates radial growth of the streamer. We describe the evolution of this streamer in terms of a cylindrical channel with growing radius R over time, as shown in figure \ref{fig:pore_model}. The growth rate will depend on the net flux of particles that are eroded off the network into the channel and swept upwards with the back flow.

\begin{table}
  \begin{center}
\def~{\hphantom{0}}
  \begin{tabular}{rlcccc}
Volume fraction & $\phi$ & \\ [5pt]
Fractal dimension & $d_f$ & \\[5pt]
Attraction range & $\delta$&=&$\dfrac{\Delta}{a}$ & =&$\dfrac{\mbox{Attractive well width}}{\mbox{Particle radius}}$ \\[8pt]
Network strength & $\epsilon$&=&$\dfrac{kT}{U}$ & =&$\dfrac{\mbox{Thermal energy}}{\mbox{Interparticle bond strength}}$ \\[8pt]
Gravitational Mason number & $G$&=&$\dfrac{4\pi\Delta\rho g a^4}{3U}$ & =&$\dfrac{\mbox{Particle gravitational energy}}{\mbox{Interparticle bond strength}}$ \\
  \end{tabular}
  \caption{The five network parameters characterizing the scenario of a freely settling colloidal gel considered in the model.}
  \label{tab:groups}
  \end{center}
\end{table}

\subsection{Growth of a critical streamer}
To arrive at the radial growth rate of the streamer or channel, consider the transport process that move particles into the channel and the reverse process of attachment into the network. The evolution of the number of particles $N$ in the cylindrical channel of height $L$ and radius $R(t)$ is given by the net flux:
\begin{equation}
\frac{dN}{dt} = J_\mathrm{in}-J_\mathrm{out} = 2\pi R L (j_\mathrm{in}-j_\mathrm{out}),
\end{equation}
where $J$ and $j$ are the rate and flux per unit area of particles into the growing channel ($in$) and out into the network ($out$), respectively. To the best approximation, for tall enough channels ($L$ is large compared to the mesh size of the gel network) the density of particles in the interior of the channel will be equal to its bulk value $\phi$, so that the number of particles is related to the geometry of the channel through $N=3\pi R^2 L\phi/4\pi a^3$. When the channel spans the height of the network, $L$ is unchanged over time and the one dimensional growth model of $R(t)$ is:
\begin{equation}
\label{eq:drfirst}
\frac{dR}{dt} = \frac{4\pi a^3}{3\phi} (j_\mathrm{in}-j_\mathrm{out})=\frac{1}{n} (j_\mathrm{in}-j_\mathrm{out}),
\end{equation}
where $n$ is the bulk number density of particles.

The flux of particles per unit area into the open channel is proportional the surface number density of particles at the network-channel interface, $\tilde{n}$, and the rate of particle bond breaking, $k_\mathrm{break}$:
\begin{equation}
j_\mathrm{in} = d_1 k_\mathrm{break}\tilde{n}.
\end{equation}
Note that throughout this paper, $d_i$ for $i=\left\lbrace 1,...,6\right\rbrace$ are unknown $O(1)$ scalar constants. The boundary between the interior of the channel at bulk density $\phi$ and the percolated gel network of fractal dimension $d_f$ constitutes the 'wall' of the streamer. For a fractal structure, the surface density of particles attached to the network at this interface is $\tilde{n} = \phi^{d_f/3}\tfrac{d_2}{a^2} \left(\tfrac{R}{a}\right)^{\left(d_f/3-1\right)}$. As one would expect, for a compact structure with $d_f=3$, $\tilde{n}$ is independent of the radius. In the absence of an external force, we hypothesize that $k_\mathrm{break}$ is set by the Kramers escape time of a particle diffusing out of an attractive well of depth $U$ and width $\Delta$\citep{kramers1940brownian}:
\begin{equation}
k_\mathrm{break} = \tau_K^{-1} = \frac{D}{a^2}\left(\frac{a}{\Delta}\right)^2\frac{U}{kT} e^{- U/(d_3 kT} =  \frac{D}{a^2}\delta^{-2}\epsilon^{-1} e^{- 1/(d_3\epsilon)},
\end{equation}
where $D$ is the diffusivity of a particle of radius $a$. The coefficient $d_3$ is included to account for the fact that the particles sit in a complex energy landscape in the gel network for which the barrier between bound and unbound states could be smaller $ U $ owing to elastic stresses stored intrinsically in the network during its formation.  Due to the back flow of fluid through the channel, the particles at the interface will feel a hydrodynamic drag, due to the shear stress, $\tau$, at the channel wall. This results in a relative force $d_4 \tau a^2 /kT$ that stretches the inter-particle bonds and accelerates the breaking process.  This activated hopping leads to a higher rate of erosion. Assuming simple Poiseuille flow in the cylindrical channel the shear stress at the wall is related to the channel radius through $\tau=\left|\nabla p\right|R/2$ and $k_\mathrm{break}$ becomes:
\begin{equation}
k_\mathrm{break} = \frac{D}{a^2}\delta^{-2}\epsilon^{-1} e^{- 1/d_3\epsilon + d_4\left|\nabla p\right|R a^2 \Delta/2kT} = \frac{D}{a^2}\delta^{-2}\epsilon^{-1} e^{- 1/d_3\epsilon + R/R^*},
\end{equation}
$R^*$ is an effective gravitational length, i.e. the characteristic length scale of the erosion process:
\begin{equation}
\label{eq:rstar}
R^*=\frac{2kT}{d_4 a^2 \Delta  \left|\nabla p\right|} = \frac{8\pi a}{3d_4}\epsilon\delta^{-1}G^{-1}.
\end{equation} 
It sets the characteristic pore size, beyond which the activated hopping dominates the bond breaking. The erosive flux of particles into the channel becomes:
\begin{equation}
j_\mathrm{in} = \frac{4\pi}{3} d_1 d_2 \phi^{d_f/3-1} R\left(\frac{R}{a}\right)^{\left(d_f/3-2\right)} \frac{nD}{a^2}\delta^{-2}\epsilon^{-1} e^{- 1/(d_3\epsilon) + R/R^*} .
\end{equation} 

The flux of individual particles currently in the bulk of the channel back onto the network is driven by diffusion according to:
\begin{equation}
j_\mathrm{out} = \frac{nD}{x} = \frac{nD}{R} f(\Pen),
\end{equation}
where $\Pen$ is the advective P\'eclet number near the wall $\Pen=\tau R^2/\eta D=\tfrac{3\pi R^3 a\left|\nabla p\right|}{kT}=\tfrac{9}{4} \epsilon G \left(\tfrac{R}{a}\right)^3$. Here, $x$ is the thickness of the diffusive boundary layer at the wall of the channel: $f(\Pen)=1$ for $\Pen\ll 1$ and $f(\Pen)\sim \Pen^{1/2}$ when $\Pen\gg 1$\citep{acrivos1965asymptotic, goddard1966asymptotic}. Except for very early times when the channel initially forms, the radius of the channel exceeds the primary particle size, $R\gg a$, and $\Pen\gg 1$ so that to first order:
\begin{equation}
j_\mathrm{out} = d_5 R\frac{nD}{a^2}  \epsilon^{1/2} G^{1/2} \left(\frac{R}{a}\right)^{-1/2}.
\end{equation}

Note that $j_\mathrm{in}\gg j_\mathrm{out}$ for $R\gg a$ so that to first approximation the flux of particles from the channel onto the network can be neglected, i.e. $j_\mathrm{out}\approx 0$. Rescaling time $t$ on the pure diffusive Kramers escape time, $\hat{t} =tD/\left(\Delta^2\epsilon\right) e^{-1/(d_3\epsilon)}=t/\tau_K$, the model predicts the following evolution equation for $R(t)$:
\begin{equation}
\frac{dR}{d\hat{t}} = \frac{4\pi}{3} d_1 d_2 a\phi^{d_f/3-1} \left(\frac{R}{a}\right)^{d_f/3-1}e^{R/R^*}.
\label{eq:ode} 
\end{equation}

There are two competing effects that set the erosion of particles and the evolution of the radius of the channel. For fractal structures with $d_f<3$, the quantity $(R/a)^{d_f/3-1} $ suggests that the growth rate of $R$ is slowed as $ R $ becomes larger as the total number of particles eroded by the back flow is decreasing \textit{relative} to the number of particles in the channel. In contrast, the term $ e^{R/R^*} $, arises because the rate of bond breaking is exponentially dependent on the shear stress exerted by back flow, which itself is linear in $R$. Hence the activated rate of erosion grows exponentially with the channel radius.  The result is that for $ R>R^* $, the erosion process accelerates exponentially. More generally, such first order ordinary differential equations, which have a super-linear flux or grow rate, exhibit a finite-time blowup singularity\citep{ide2002oscillatory}. 

In the case of this erosion model, we predict an exponentially dependent flux, an ultrafast type of super-linear growth, so that the channel radius will grow infinitely large at a critical point in time. This blowup time is defined such that: $R\rightarrow\infty$ as $t\rightarrow t_\mathrm{blowup}$. This is the main result of our phenomenological model. At a certain critical point in time in freely settling gels a hydrodynamic instability occurs. The channel radius grows unstably to span the width of the gel.  In practice, this corresponds to the streamer being of comparable size to the width of the container, as seen by \citet{starrs2002collapse}, when large scale fluid back flow deconstructs the gel network. The solid is ripped apart and the gel rapidly collapses. Note, that due to the exponential growth profile, the fate of the network is determined long before $R$ reaches the system size. When $R\geq R^*$, the channel growth rate is exponential and catastrophic collapse is unavoidable. However, macroscopically the radial growth of the channel may not manifest itself in a dramatic increase in the settling velocity of the network's top interface until $t$ is right near $t_\mathrm{blowup}$. 

To explain this observation consider that for an incompressible Newtonian suspending fluid the settling velocity of a gel and therefore the velocity of the interface will be proportional to the average fluid back flow velocity, $\left\langle u_{f} \right\rangle$, driven by the constant pressure gradient $|\nabla p|$ across the network. Here, $\left\langle u_{f} \right\rangle$ is the average volumetric flow rate of fluid back flow through the intact gel \emph{and} the streamer channel. For simplicity, the intact gel network can be thought of as a porous Darcy medium with permeability $\kappa$, whereas the streamer is a cylindrical channel with Poiseuille flow, as before.

The fluid velocity through the gel network is given by:
\begin{equation}
u_f^\mathrm{Darcy} = -\frac{\kappa}{\eta}\nabla p.
\end{equation}
The total flow rate through the medium with cross sectional dimension $L$ and a cylindrical pore of radius $R$ within it is $Q_f^\mathrm{Darcy} = -\left(L^2-\pi R^2\right)\frac{\kappa}{\eta}\nabla p$. In the cylindrical channel the velocity profile is: $u_f^\mathrm{Channel} = -\tfrac{1}{4\eta}\nabla p\left(C-r^2\right)$, where $r$ is the radial position and the constant $C$ is set by the boundary condition at the network-channel interface, $r=R(t)$. This boundary condition will be highly dependent on the exact fractal structure of the network.  At lowest order this can be captured by introducing a Navier slip length $\lambda$ so that at the boundary\citep{beavers1967boundary}: $u_f^\mathrm{Channel}\left(r=R(t)\right)-u_f^\mathrm{Darcy}=-\lambda \frac{du_f^P}{dr}$. Therefore the resulting parabolic flow profile inside the channel is:
\begin{equation}
u_f^\mathrm{Channel} = -\frac{1}{4\eta}\nabla p\left(R^2-r^2+2\lambda R + 4\kappa\right),
\end{equation}
with a volumetric flow rate: $Q_f^\mathrm{Channel} = \tfrac{-\pi R^2}{8\eta}\nabla p\left(R^2 + 4\lambda R + 8\kappa\right)$. Thus the average fluid back flow velocity over the pore and porous medium as the gel settles is given by:
\begin{equation}
\label{eq:fluidvel}
\left\langle u_{f} \right\rangle = \frac{Q_f^\mathrm{Channel} + Q_f^\mathrm{Darcy}}{L^2} = -\frac{1}{\eta}\nabla p\left[\kappa +\frac{\pi}{8}\left(\frac{R}{L}\right)^2\left(R^2+4\lambda R\right)\right]
\end{equation}

The channel size will be negligible compared to the overall size of the porous medium for the majority of the growth process, i.e. $R/L\ll 1$. The rate of collapse of the gel is therefore limited by the back flow of the fluid through the solid network. It is controlled by $\kappa$, yielding a uniform settling profile of the top interface as if there were no growing channel\citep{manley2005gravitational}. However, as the settling proceeds and $t\rightarrow t_\mathrm{blowup}$, $R$ will grow rapidly and the channel cross section will become significant, $R/L\sim O(1)$. The second term in equation  \eqref{eq:fluidvel} will dominate the fluid back flow velocity. Consequently, the interface velocity will appear to increase rapidly as the size of the channel becomes comparable to the size of the gel network. Therefore $t_\mathrm{blowup}$ sets the timescale for the onset of macroscopic collapse. 

\subsection{Finite-time singularity in channel growth}\label{sec:finite}
A closed form expression for the finite-time singularity, or blowup time predicted by the model can be found by specifying the initial value for the radius of the channel, $R(\hat{t}=0)$:
\begin{equation}
\label{eq:full}
\hat{t}_\mathrm{blowup} = d_6 R^{*^{2-d_f/3}} \Gamma(2-d_f/3,R(0)/R^*) \phi^{1-d_f/3}, 
\end{equation}
where $\Gamma(a,x)=\int_x^\infty s^{a-1}e^{-s} ds$ is the incomplete Gamma function and $d_6=\left(d_1 d_2 4\pi/3\right)^{-1}$. Our model predicts that the finite-time singularity is intrinsic to all freely settling colloidal gel networks.  One interesting feature is that for an initial channel radius of zero, $ R(0) = 0 $, the blowup time is non-zero.  In fact, all initial channel radii $ R(0) \ll R^* $ have nearly the same blowup time: $ \hat{t}_\mathrm{blowup} = d_6 R^{*^{2-d_f/3}} \Gamma( 2- d_f/3, 0 ) \phi^{1-d_f/3}  \approx 0.9  d_6 R^{*^{2-d_f/3}} \phi^{1-d_f/3} $, when $ d_f = 2 $.  In essence, the gel is filled with pores having a heterogeneous size distribution.  The fluid will select the largest initial pore and erosion of that pore will be favoured over others.

As discussed in $\S$\ref{sec:future} heterogeneities within the initial gel may form from other processes beyond mere thermally driven restructuring\citep{lu2008gelation}.  These can have a number of origins, including falling debris that accumulates at an interface\citep{harich2016gravitational}, external forcing fields\citep{teece2014gels}, or included bubbles. Once a streamer is born with initial size $ R(0) $, the growth rate is universally described by \eqref{eq:ode} and the final stages of breakup leading to collapse are identically captured by the model with $R^*$ being the only controlling parameter. 
 
Because the incomplete Gamma function takes on values $[0,1]$, the behavior of the blowup time is dictated by the network parameters in the model.  When $ R(0) = 0 $:
\begin{equation}
\label{eq:scaling}
t_\mathrm{blowup}\sim \tau_D \phi^{1-d_f/3}\delta^{d_f/3} \epsilon^{3-d_f/3} e^{1/d_3\epsilon}G^{d_f/3-2},
\end{equation}
where the only unknowns are the scalar coefficient of proportionality and the coefficient associated with Kramers hopping process, $d_3$. This is a clear prediction of how the point in time where the hydrodynamic instability and collapse occurs relates to properties of the gel network. These are the adjustable material parameters that are available to engineer stability into the particulate network.  Before we proceed to discuss the utility of our model prediction in $\S$\ref{sec:discussion}, we test its validity using extensive computer simulations and comparisons to published experimental data in the next section.

\section{Model validation with Simulations and Experiments}
\label{sec:validation}
To assess how well the model captures the process of gravitational collapse we first compare it to observations of simulations of settling gels.  In $\S$\ref{sec:methodology} and in $\S$\ref{sec:results} we describe the simulation conditions in greater detail and present the results of the parametric sweep in terms of the network parameters introduced in $\S$\ref{sec:model}. In $\S$\ref{sec:seededhole} we show the dynamics for a range of seeded channel sizes. While good agreement between theory and simulations is necessary, to be certain about the model validity, it has to reproduce collapse dynamics observed in experiments. In $\S$\ref{sec:experiment} we present comparisons with two published experimental studies.

\subsection{Simulation Methodology} 
\label{sec:methodology}
In order to study hydrodynamic instabilities during gel collapse and observe the breakdown of the network, the effects of fluid flows and hydrodynamic forces have to be modelled accurately in large scale simulations. We have recently developed methods for rapid calculation of hydrodynamic interactions in suspensions of mono-disperse spheres\citep{swan2016rapid, fiore2017rapid}, where we use the Rotne–-Prager-–Yamakawa tensor(RPY) to account for long-ranged hydrodynamic interactions with great fidelity\citep{rotne1969variational}. The positively-split Ewald (PSE) algorithm makes the cost of computing Brownian displacements in simulations of colloidal scale particles with hydrodynamic interactions comparable to the cost of computing deterministic displacements in freely draining simulations. The method relies on a new formulation for Ewald summation of the RPY tensor, which guarantees that the real-space and wave-space contributions to the tensor are independently symmetric and positive-definite for all possible particle configurations. Brownian displacements are drawn from a superposition of two independent stochastic samples: a wave-space (far-field) contribution, computed using techniques from fluctuating hydrodynamics and non-uniform fast Fourier transforms; and a real-space contribution, computed using a Krylov subspace method. The combined computational complexity of drawing these two independent samples scales linearly with the number of particles  enabling hydrodynamic simulations with system sizes up to $10^6$ particles\citep{fiore2017rapid}.

Extensive simulations of freely settling, attractive, hydrodynamically-interacting, colloidal particles of size $a$ in a solvent of viscosity $\eta$ are performed. The simulations contain $N_{sim} = 216,000$ particles having a volume fraction $\phi$ in a cubic simulation box of length $L_{sim}$ with periodic boundary conditions. $N_{sim}$ has been selected to avoid any system size effects\citep{varga2015hydrodynamics} and to be able to resolve large scale structural changes\citep{varga2018large}. Other choices of aspect ratio, including stretched and flattened gel columns, have been investigated and do not affect the simulation results that follow. Any interactions with the container and other potential wall effects are neglected. While the interactions with walls may influence collapse\citep{poon2002physics} the onset of the hydrodynamic instability occurs far from any boundaries in the feely settling regions of gels and effects of the sample geometry may be insignificant\citep{secchi2014time}. Zero volume flux boundary conditions on the simulation box ensure sedimentation models the free falling zone\citep{buscall1987consolidation}, where the gel network is freely settling. Note that as a consequence, both the bottom of the container and the compacting cake region, shown previously to play no role in collapse, are ignored. Furthermore the processes at the top surface of the gel and the role of the meniscus in inducing collapse are not under study. The simulation can be viewed as modelling micron sized colloids and a millimetre sized gel cross section deep inside a sedimenting network that is in a state of free fall.

We introduce a short-ranged attraction, mimicking the polymer induced depletion attraction in experimental systems\citep{russel1989colloidal, Poon1997, Lu2006}, and model it through an Asakura-Oosawa form\citep{Asakura1958}:
\begin{equation}
U_{A}(r) = -U \frac{2 ( 2 a (1+\delta) )^3 - 3 r (2 a ( 1 + \delta ) )^2 + r^3}{2 ( 2a (1+\delta) )^3-6a ( 2a (1+\delta) )^2+(2a)^3},
\end{equation}
for particle separations $ r$ in the range of  $ 2 a < r < 2 ( a + R_g ) $. The polymer radius of gyration, $R_g$, relative to the colloid particle size is varied, $\delta=R_g/a$, and the pairwise depletion strength at contact sets the network strength, $\epsilon = kT / U$, from the athermal limit, $\epsilon=0$, to a hard-sphere dispersion, $\epsilon\rightarrow \infty$. The Heyes-Melrose potential-free algorithm ensures hard-sphere repulsion at contact\citep{Heyes1993}. The uniform gravitational load on all particles is tuned through the gravitational Mason number, $G$. We study the settling systems over a range of attraction ranges: $\delta = 0.075 - 0.15$, strengths: $\epsilon = 0.01 - 0.2$, Mason numbers: $G = 0.1 - 1.0$, and volume fractions:  $\phi = 5-50\%$(see table \ref{tab:groups} for definitions of all dimensionless quantities). Initially, the dispersion is allowed to gel for 500 bare diffusion steps, $\tau_D = 6\pi\eta a^3 / kT$ in the absence of gravity. Use of the box counting method determines the Minkowski-Bouligand dimension\citep{falconer2004fractal}, $d_f$, for each gel, which characterizes the meso-scale structure and tortuous nature of the random network. Note that due to the finite system size and finite particle size the gels are not true fractal objects over all length scales. At time $t=0$ we introduce a finite gravitational Mason number $G$ and observe the process of free settling over a simulation time of $2500 G/\epsilon \tau_D = 2500\tau_G$ , where $\tau_G$ is the characteristic settling time, the time it takes a single particle to settle its own radius in bulk fluid. As the network moves downward, fluid back flow, particle erosion and the eventual failure of the network are observed. All reported simulation results are averaged over 3 independently generated samples for each combination of $\delta$, $\epsilon$, $G$ and $\phi$. 

\subsection{Comparison with simulations of freely settling colloidal gels}
\label{sec:results}
The colloidal gel networks in the simulations all exhibit gravitational collapse during free settling. While the exact time point at which this occurs strongly depends on the network parameters $\delta$, $\epsilon$, $G$ and $\phi$, all gels eventually experience a hydrodynamic instability and fail catastrophically as shown in figure \ref{fig:pore_simulation} (see the supplementary materials for further videos of collapse). Initially, the network settles uniformly under its own weight and fluid flows upwards through the gel pores. After an initiation period, a single streamer nucleates in the gel. This streamer grows radially as individual particles and small clusters are eroded and swept upwards with the back flow. The streamer then grows and spans the height of the settling gel. There now exists a continuous channel for fluid back flow and the entire network is destabilized as the streamer rapidly expands in the radial direction. Eventually, portions of the gel compact, network integrity loss occurs and large domains of the gel move independently both up and downwards. The gel is no more and the network has undergone catastrophic collapse.
\begin{figure}
  \centerline{\includegraphics[width=0.9\columnwidth]{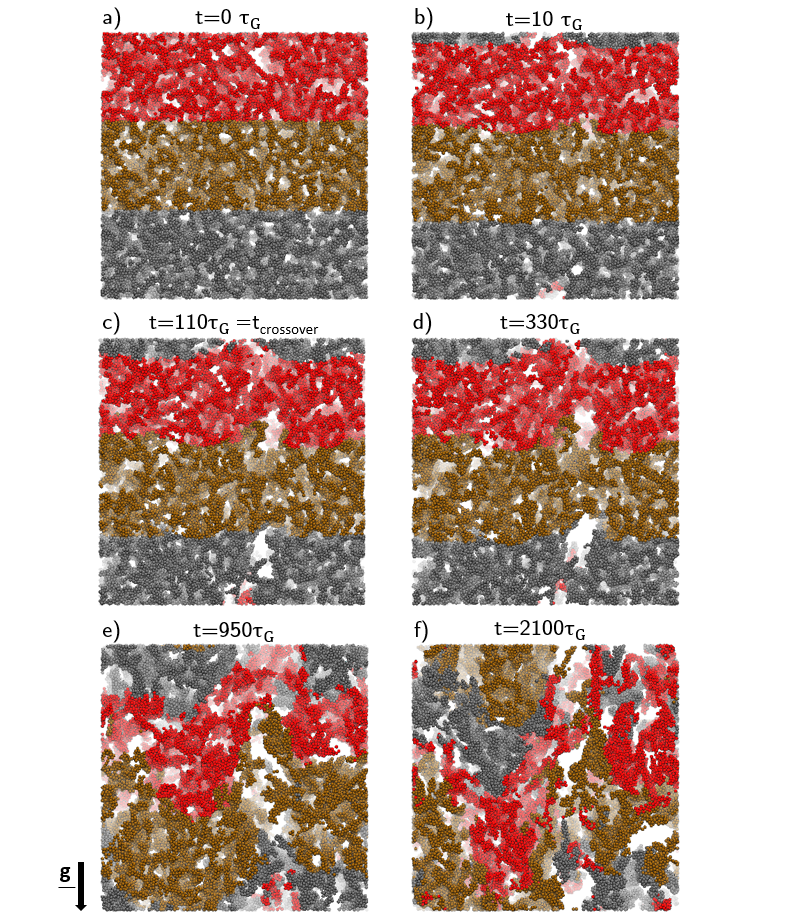}}
  \caption{In simulations we observe freely sedimenting gels in a gravitational field \textbf{g} and study the effects of fluid back flow over time (see $\S$\ref{sec:methodology} for simulation details). These are snapshots of a simulation with $\delta=0.1$, $\epsilon=0.05$, $G=0.5$ and $\phi=20\%$ (see supplementary materials for a movie). The differently coloured layers (colour online) are purely for illustrative purposes, indicate initial particle positions in the gel, and are meant to guide the eye through the breakdown of the network during free settling. The dispersion gelled over $500\tau_D$ in the absence of gravity and has a structure characterized by $d_f=2.05$. a) At $t=0\tau_D$ gravity is turned on in the simulation and the network begins to settle. b) After initial uniform settling, a single cylindrical streamer nucleates (bottom center). c) The streamer is both growing radially and its height spans the bottom half of the network. d) At the onset of settling rate increase ($t=t_\mathrm{crossover}$) the streamer spans the height of the gel. e) The streamer continues to grow radially, filling the entire sample, destabilizing and changing the uniform settling of the network. f) There is complete loss of network integrity as entire aggregates are ripped off the gel.\label{fig:pore_simulation} }
\end{figure}

\begin{figure}
  \centerline{\includegraphics[width=0.75\textwidth]{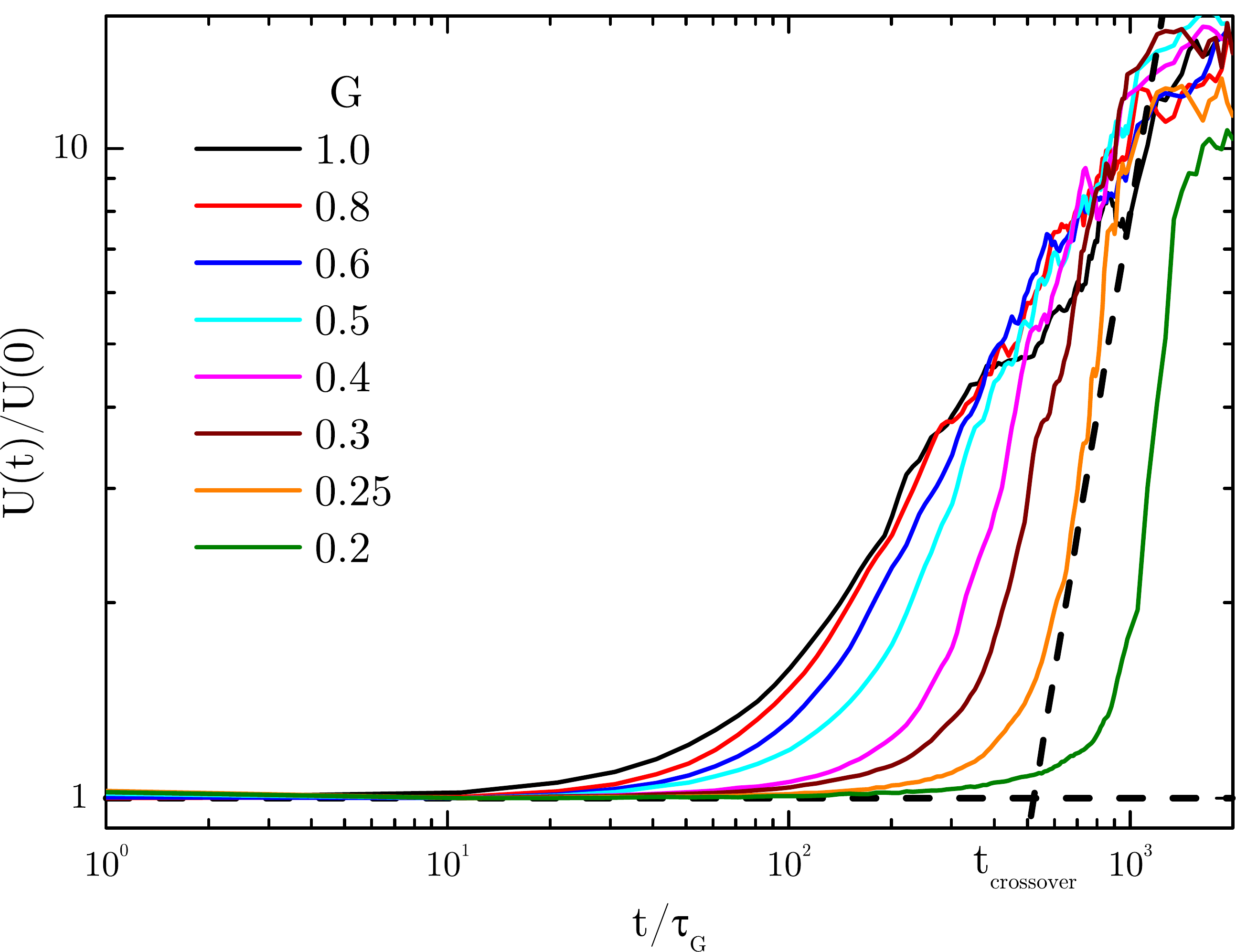}}
  \caption{The average network settling rate $U(t)$ normalized by the initial rate $U(0)$ as a function of simulation time in units of bare diffusion time $\tau_D$ for increasing $G$ with $\delta=0.1$, $\epsilon=0.05$ and $\phi=20\%$. The onset of rapid growth in settling rate, $t_\mathrm{crossover}$, is the point in time where the power-law growth at a given $G$ intersects the plateau of uniform settling $U(t)=U(0)$, marked by the two dashed lines.\label{fig:vel_withlabel}}
\end{figure}

We quantify this process using two independent metrics, the evolution of the network settling velocity and the growth of the streamer volume over time. The average network settling velocity, $U(t)$, is measured by computing the velocity of the centre of mass of the gel in the frame of an external observer, the lab frame as opposed to the frame of zero volume flux, the Lagrangian perspective, for all gels under study (note that we exclude the velocity of particles that are not attached to the percolated structure). Figure \ref{fig:vel_withlabel} shows this network settling velocity normalized on its initial value, $U(0)$, as a function of dimensionless simulation time for increasing values of $G$ and constant $\delta$, $\epsilon$ and $\phi$. The data exhibit three distinct regimes of settling. As seen, the network begins to settle with a constant initial uniform velocity. Then at a certain point, the velocity grows as a power-law in time until it enters the third regime, where it reaches a new plateau of the settling velocity with $U_\mathrm{final}/U(0)\sim O(10)$, consistent across all gels studied. Interestingly, \citet{starrs2002collapse} also observed a ten-fold increase in their velocities from initial settling to final collapse. 

In agreement with visual observations, a stronger gravitational force results in an earlier onset of the power-law growth in settling velocity. We term the transition to power-law growth the crossover time, $t_\mathrm{crossover}$, and identify it in each simulation as the point where the best-fit power-law line intersects the initial settling velocity, as illustrated in figure \ref{fig:vel_withlabel} by the dashed lines. $t_\mathrm{crossover}$ changes by orders of magnitude depending on the values of the network parameters, but eventually a transition to an increasing settling velocity and final plateau is observed for even the strongest gels studied here. 

In parallel with the measurements of the network settling velocity, we track the growth of the overall streamer volume for each gel, employing an approach similar to the box counting method.  The density of particles in each box is counted and the corresponding distribution of particle densities is computed. For a nascent randomly percolated gel structure of fractal dimension $d_f$, this distribution should resemble a Gaussian with a mean related to the bulk volume fraction $\phi$ and the width of the curve a function of $d_f$.   The difference between the observed distribution and the best-fit Gaussian distribution of intact network pore volume yields an estimate for the volume of the streamer. 

\begin{figure}
\begin{subfigure}[b]{0.48\textwidth}               
\includegraphics[width=\linewidth]{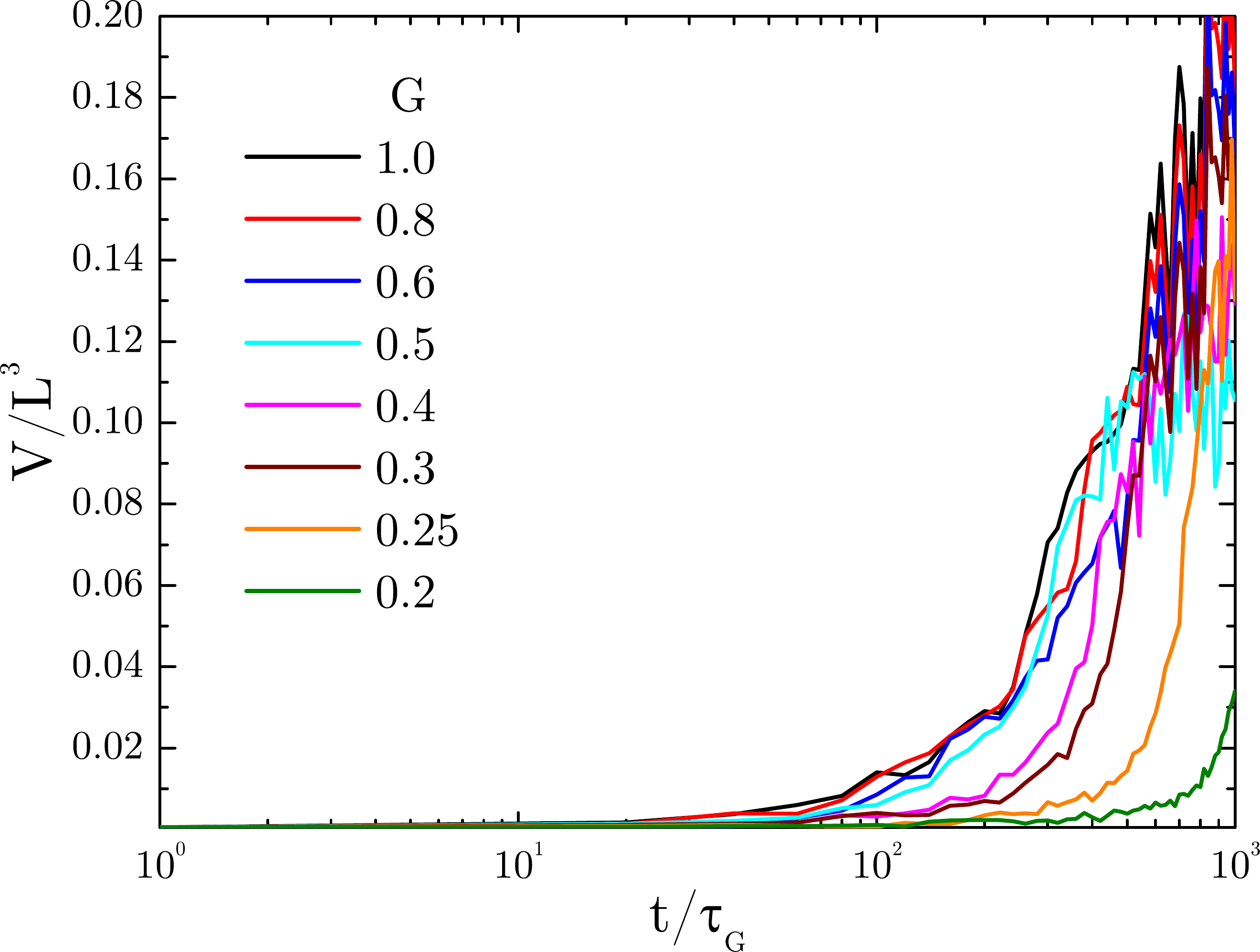}                
                \caption{The growth of streamer volume normalized on the simulation box volume plotted as a function of simulation time for increasing $G$ with $\delta=0.1$, $\epsilon=0.05$ and $\phi=20\%$.\label{fig:poreV}}
\end{subfigure}\hfill
\begin{subfigure}[b]{0.48\textwidth}
\includegraphics[width=\linewidth]{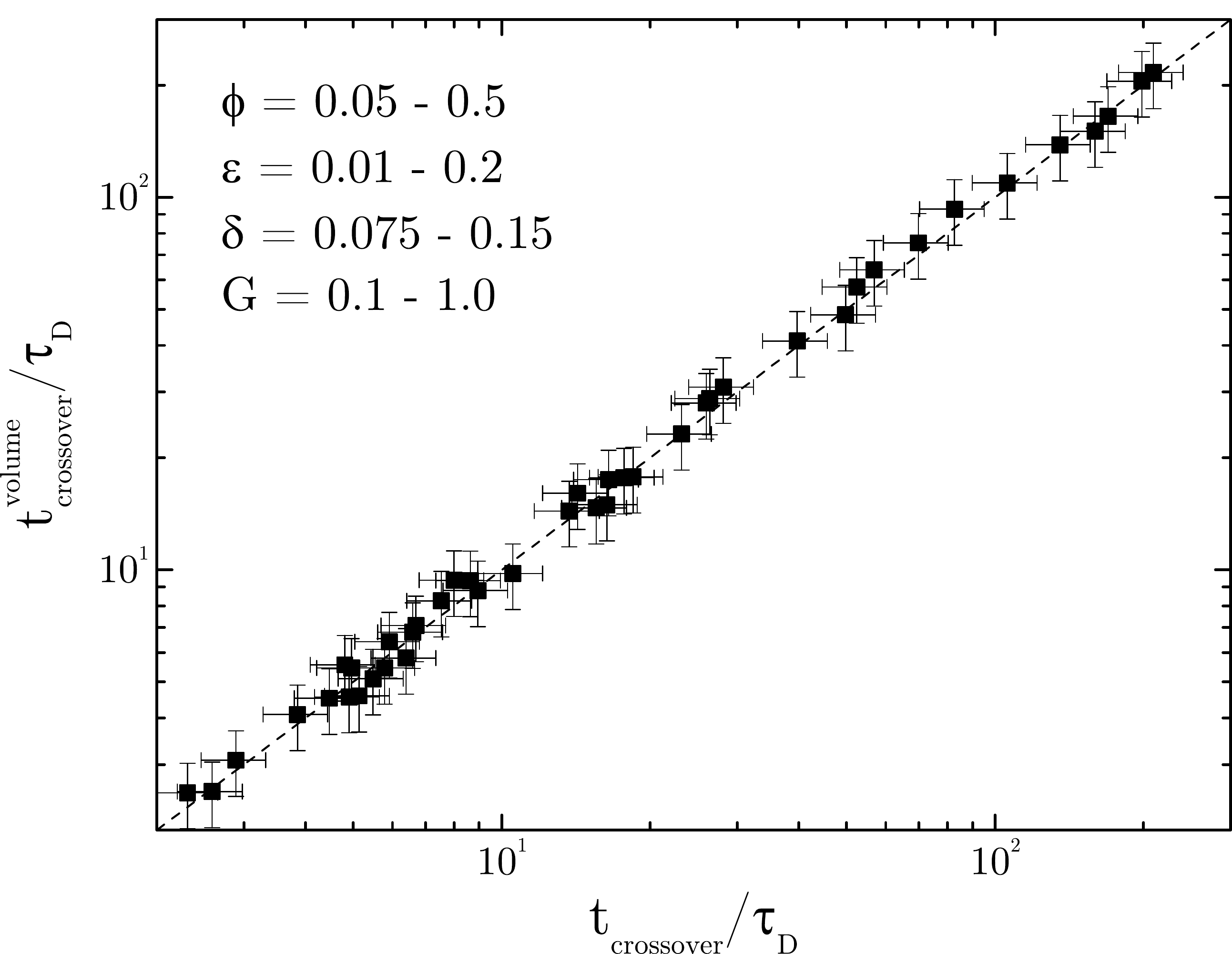}
                \caption{The onset of rapid growth in streamer volume $t_\mathrm{crossover}^{volume}$ plotted vs. the onset of  power-law growth in network settling velocity, $t_\mathrm{crossover}$.\label{fig:poreparity}}
\end{subfigure}
\caption{There is a direct correlation between the increase in streamer volume and the accelerating settling network, which supports the model's premise that streamer nucleation and growth is a cause for loss of network integrity in settling gels.\label{fig:volume}} 
\end{figure}

Figure \ref{fig:poreV} plots the evolution of the streamer volume $V(t)$ normalized on the simulation box volume for the same set of gels as in figure \ref{fig:vel_withlabel}. Initially there is no noticeable streamer present, i.e. its volume is below the sensitivity threshold of the measurement technique. At a certain point in time, which decreases with increasing G, $V(t)$ exhibits a power-law growth, very similar to the behaviour exhibited by $U(t)$. Again, it is possible to extract a crossover time, $t_\mathrm{crossover}^{volume}$, using the same method as before. In figure \ref{fig:poreparity} we plot $t_\mathrm{crossover}^{volume}$ vs. $t_\mathrm{crossover}$ and find that the two timescales are identical to within the measurement errors. The acceleration of the settling network and its ultimate collapse is directly correlated with the nucleation and subsequent growth of the streamer in the gel in accordance with the  mechanism described by the model. 

\begin{figure}
        \begin{subfigure}[b]{0.49\textwidth}
                \includegraphics[width=\linewidth]{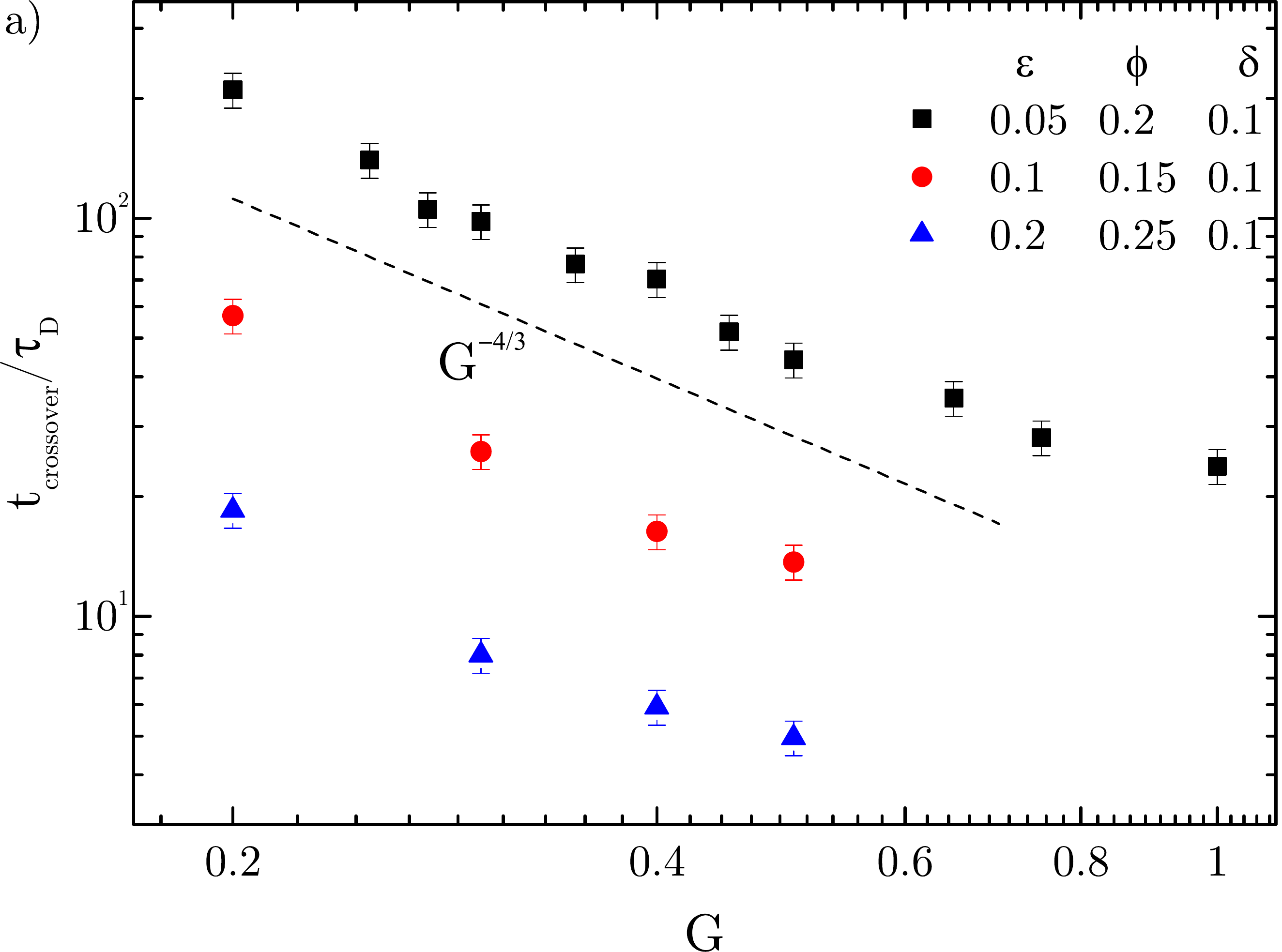}
                \caption{\label{fig:G}  }            
        \end{subfigure}
        \hfill
        \begin{subfigure}[b]{0.49\textwidth}
                \includegraphics[width=\linewidth]{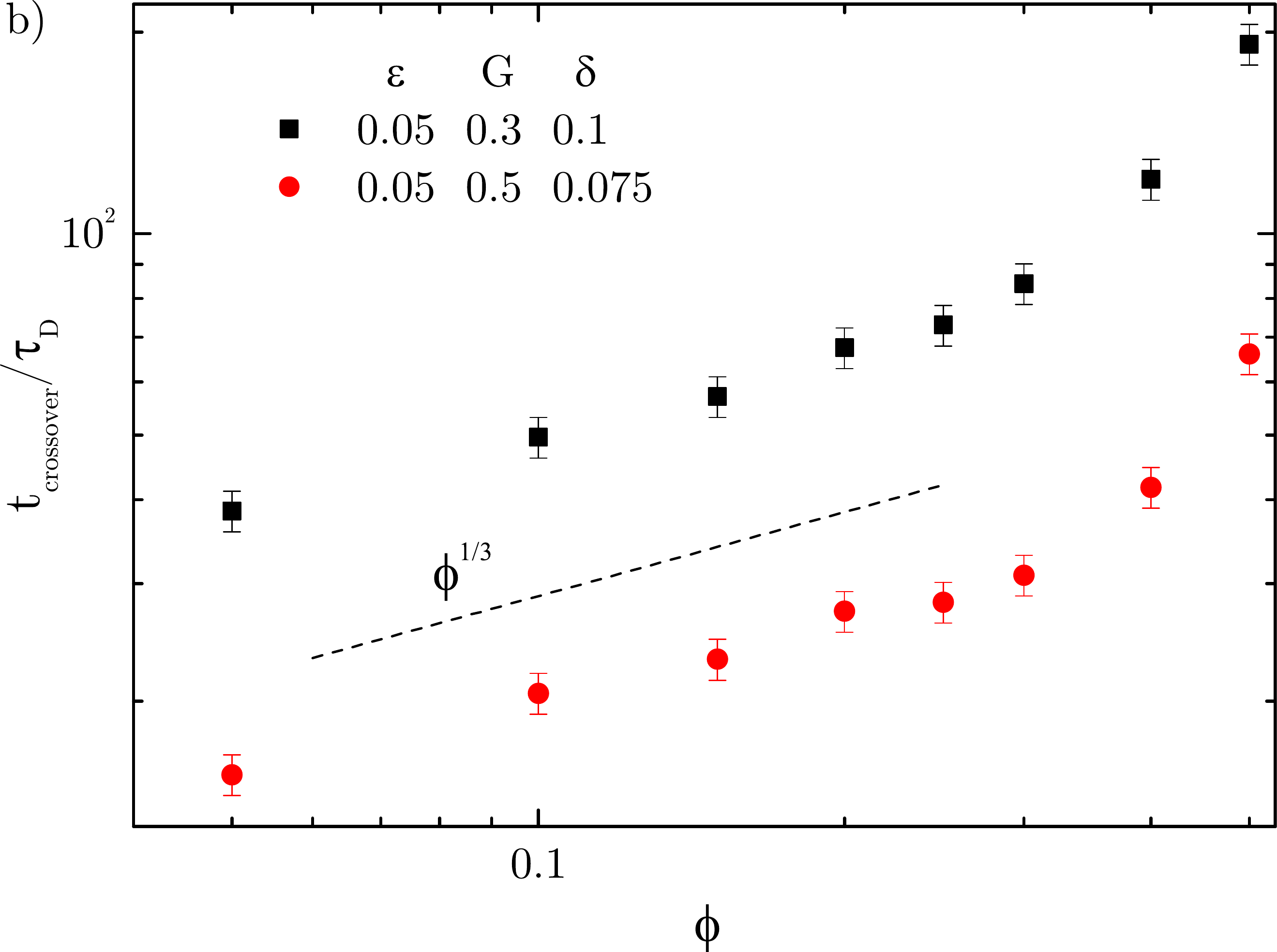}
                 \caption{\label{fig:phi}}               
        \end{subfigure}
        \vskip -0.5em
        \begin{subfigure}[b]{0.49\textwidth}
                \includegraphics[width=\linewidth]{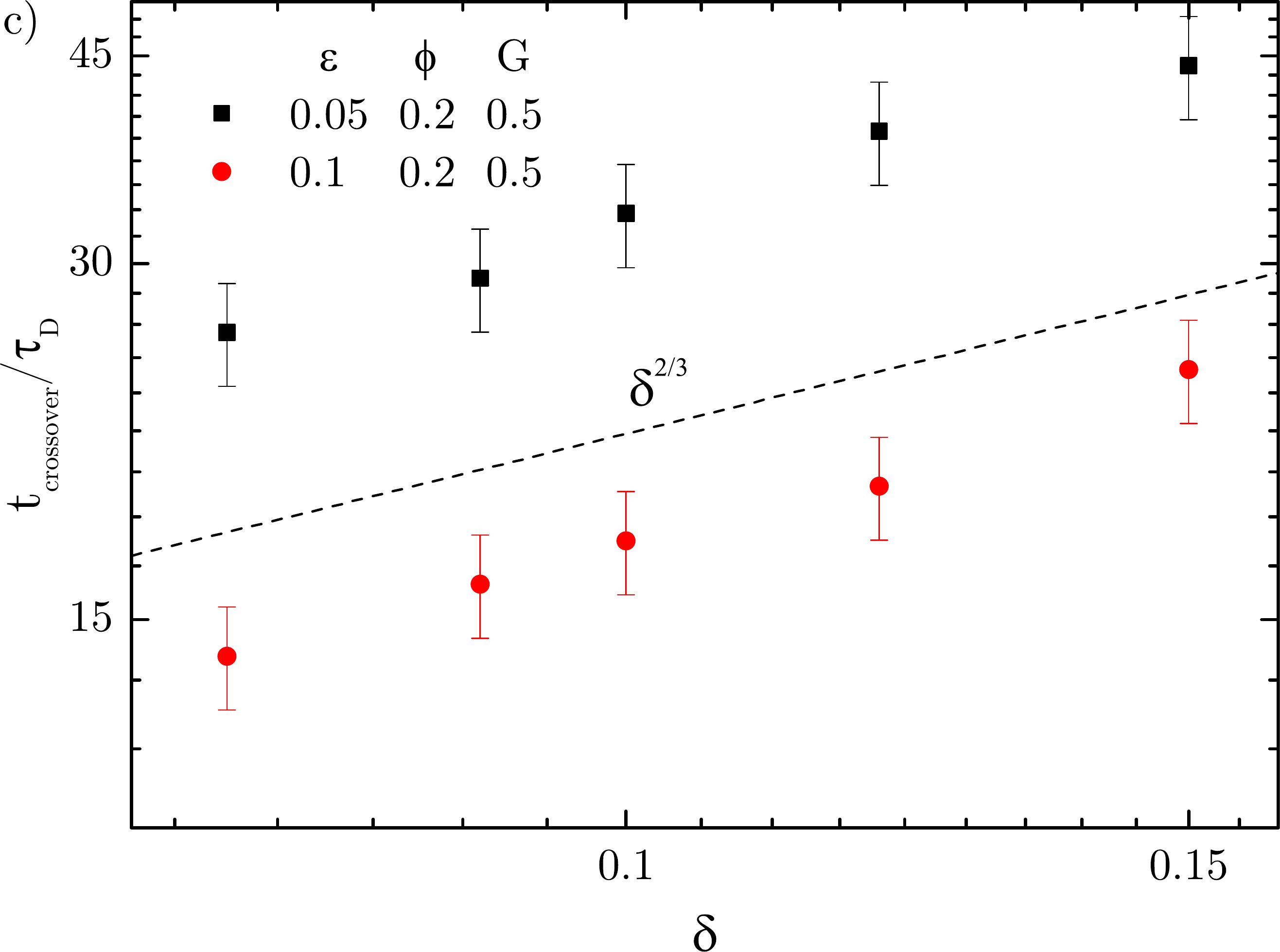}
                \caption{\label{fig:delta}}
        \end{subfigure}
        \hfill
        \begin{subfigure}[b]{0.49\textwidth}
                \includegraphics[width=\linewidth]{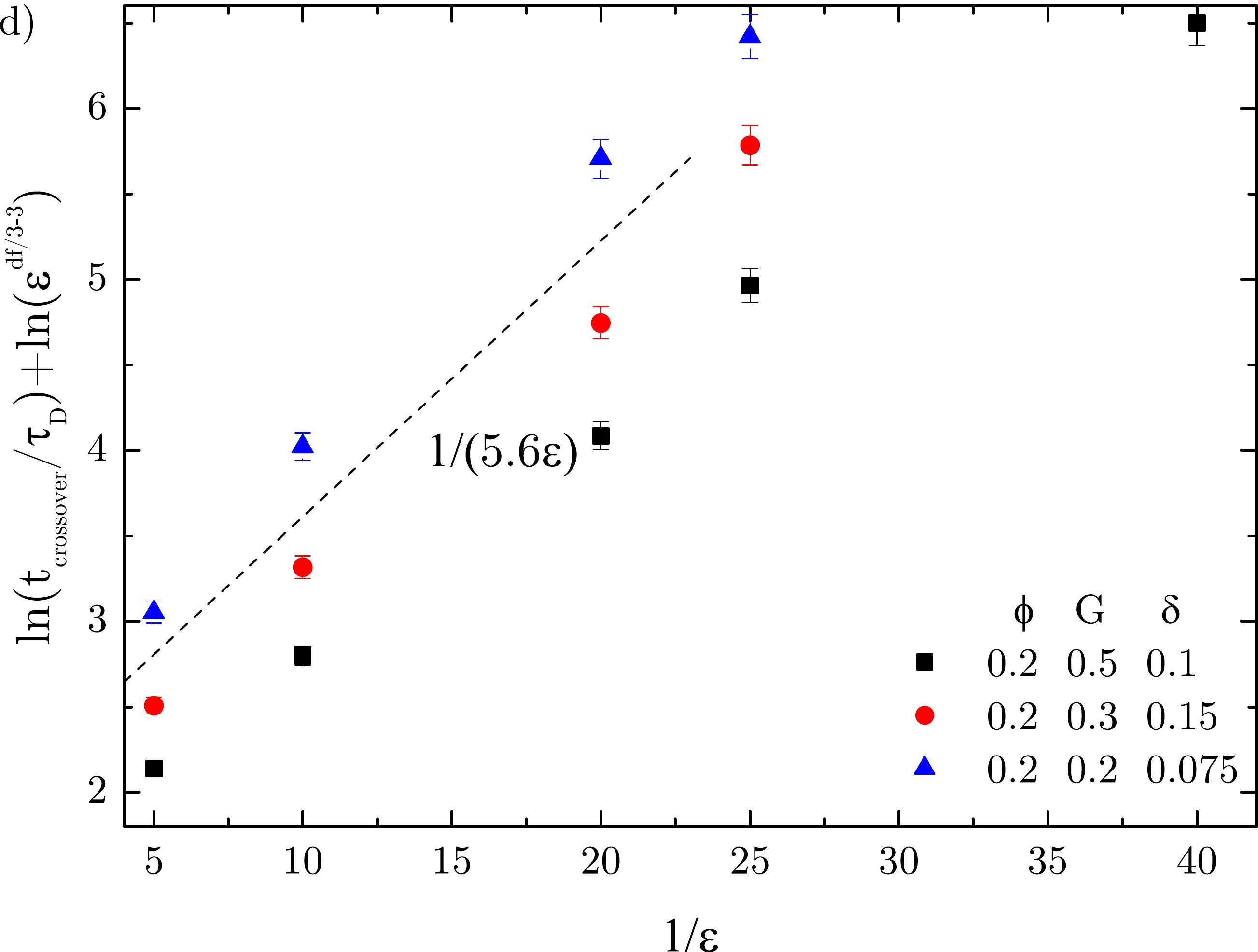}
                \caption{\label{fig:eps}}
        \end{subfigure}%
        \caption{The onset of rapid network collapse, $t_\mathrm{crossover}$ as measured in dynamic simulations is plotted individually as a function of the network parameters $G$, $\phi$, $\delta$ and $\epsilon$ respectively, while all others are held constant. The collapse dynamics exhibit the power-law behaviour predicted by the micromechanical model for networks with $d_f\approx 2$. Each data point is the average of three independent simulations and error bars represent $95\%$ confidence intervals. The slope of the best-fit line in $d)$ yields the scaling coefficient $d_3$ that sets the relevant strength of attraction, $d_3\approx 5.6$. \label{fig:details}}      
\end{figure}

The crossover time measured in simulations marks the beginning of the collapse of the gel, after which the network rapidly loses its integrity. While the dynamic simulations exhibit the same qualitative behaviour as described by the model, it remains a question whether \eqref{eq:scaling} can predict how $t_\mathrm{crossover}$ depends on the network parameters. Figure \ref{fig:details} examines the dependence of $t_\mathrm{crossover}$ on $G$, $\phi$, $\delta$ and $\epsilon$ separately. 

Figure \ref{fig:G} shows the effect of increasing gravitational Mason number for three different network strengths and volume fractions. Also shown is the expected scaling behaviour as given by \eqref{eq:scaling} for a network with fractal dimension $d_f=2$: $t_\mathrm{blowup}\sim G^{d_f/3-2}=G^{-4/3}$. Indeed for a range of gravitational Mason numbers for all three conditions the crossover time exhibits the scaling that the model would predict. Small deviations from the $-4/3$ scaling are expected as the measured $d_f$ for the gels in these simulations range between $1.7-2.3$. For large gravitational forces ($G> 1$) a different mechanism controls the network collapse. $t_\mathrm{crossover}$ appears independent of the gravitational Mason number and the situation is one of weakly aggregated clusters settling freely\citep{huh2007microscopic}.

Next the dependence of $t_\mathrm{crossover}$ on the volume fraction is analysed for two different combinations of $\delta$, $\epsilon$ and $G$ along with the expected scaling behaviour from the model, assuming $d_f=2$: $t_\mathrm{blowup}\sim \phi^{1/3}$. The crossover time exhibits roughly the model behaviour for a large range of volume fractions, as shown in figure \ref{fig:phi}. As the fractal dimension is especially sensitive to $\phi$, it is no surprise that  $t_\mathrm{crossover}$ deviates slightly from the $\phi^{1/3}$ scaling. The model prediction breaks down for $\phi>30\%$, at which point the starting material cannot be considered a gel, but instead approaches the properties of a colloidal glass. Clearly, other restructuring processes dominate settling in these dense structures\citep{zaccarelli2007colloidal}. 

Similarly, the critical time exhibits the model predicted dependence on the attraction range, $t_\mathrm{crossover}\sim\delta^{d_f/3}$ as shown in figure \ref{fig:delta}. For the values of $\delta$ investigated here, the law of corresponding states suggests that the thermodynamic restoring forces for these dispersions will all be vary similar\citep{noro2000extended}. To maintain the short-range nature of the colloidal bonds however, $\delta$ could not be increased above this narrow range since it known that the evolution of the collapse dynamics are markedly different for gels with long-range attractions \citep{teece2011ageing}. So, agreement with the model should be considered tentative.

Finally, we consider the effect of the network strength $\epsilon$ on $t_\mathrm{crossover}$. Since \eqref{eq:scaling} suggests both a power-law and exponential dependence on $\epsilon$, figure \ref{fig:eps} plots $\log \left(\epsilon^{d_f/3-3} t_\mathrm{crossover}/\tau_D\right)$ vs. $\epsilon^{-1}$. Note, in the model, the proportionality constant modulating the strength of attraction, $d_3$ is left undetermined. Indeed there appears to be a linear relation between the timescale of collapse of the network and the Kramers escape rate, as previously observed experimentally by \citet{teece2014gels}. The slope of the best-fit line for all three combinations of $\delta$, $G$ and $\phi$ indicates that the best choice for the constant is $d_3=5.6$ and the crossover time exhibits the relationship with $\epsilon$ that the model predicts. While we cannot provide a physical explanation for this particular value, it is a result of the simplified approach to approximate the network erosion in terms of the bond breaking rate or escape probability of individual particles from the attractive well. For very large attractions, $\epsilon\leq 10^{-2}$ a small deviation from the predicted scaling is observed. $\S$\ref{sec:future} discusses possible reasons for this and ways to improve the model. 

\begin{figure}
  \centerline{\includegraphics[width = 0.75\textwidth]{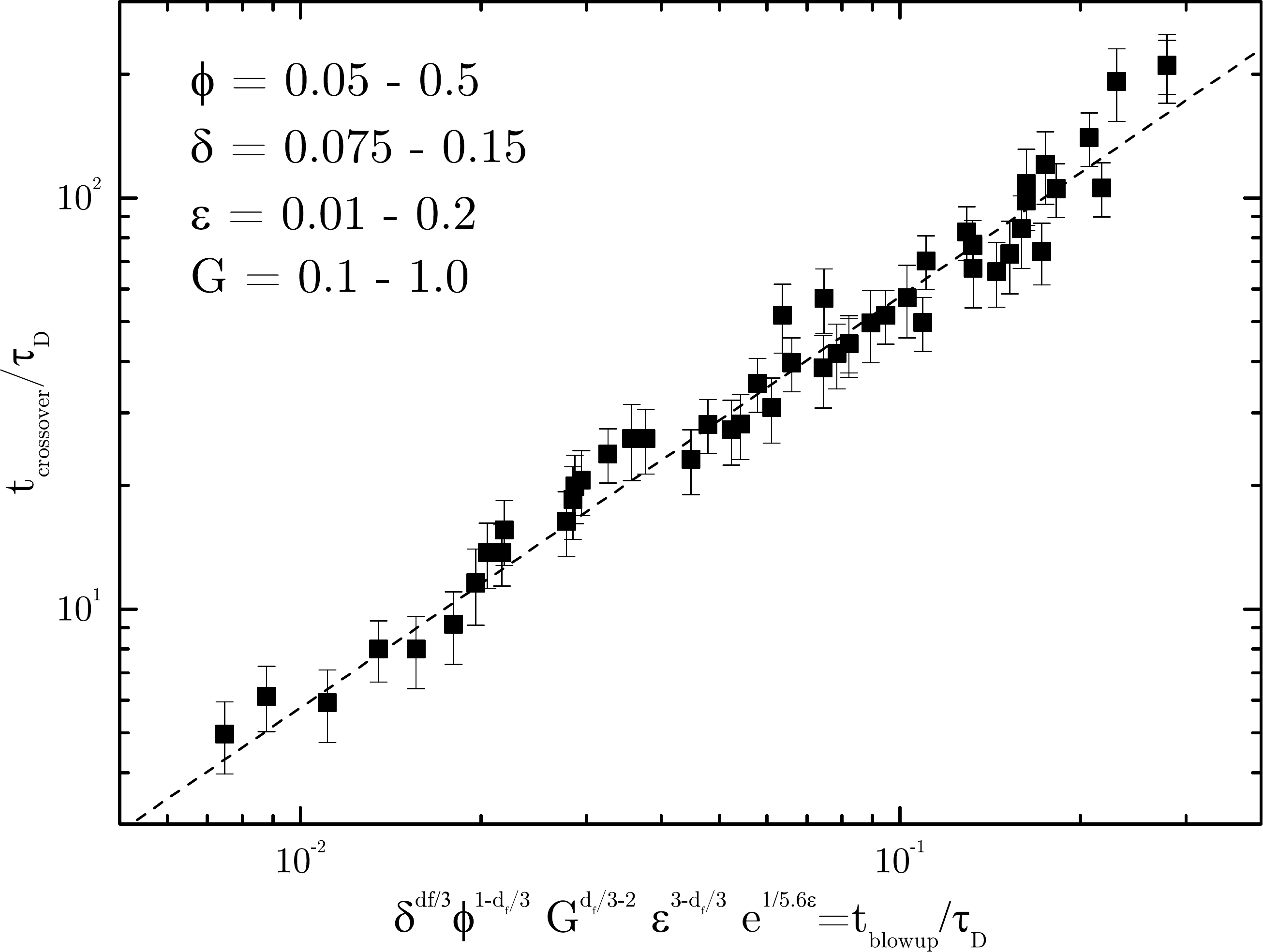}}
  \caption{The onset of rapid network collapse, $t_\mathrm{crossover}$ as measured in dynamic simulations is plotted as a function of the model prediction $t_\mathrm{blowup}$, computed for the combination of  $\delta$, $\epsilon$, $G$, $\phi$ and $d_f$ used in the respective simulation run. Each data point is the average of three independent simulations with the same unique combination of parameters in the parameter space and error bars represent $95\%$ confidence intervals. The model predicts the collapse dynamics to within a scalar coefficient, the value of which is given by the slope of the best-fit line through the data (dashed line).\label{fig:parity}}
\end{figure}

It appears that the phenomenological model adequately predicts the dependence of the critical timescale in dynamic simulations on the dimensionless parameters and captures the essential features of collapse. Equation \eqref{eq:full} gives a quantitative prediction for the critical blowup time to within a scalar constant, provided only the values of the five dimensionless network parameters. In figure \ref{fig:parity} we plot the measured $t_\mathrm{crossover}$ vs. $t_\mathrm{blowup}$ and find a direct parity between the two quantities. Here, every data point is an individual combination of $\delta$, $\epsilon$, $G$ and $\phi$ and the calculated $d_f$, averaged over three independent simulation runs. Given this linear relationship, the slope of the best-fit line through the data can be used to obtain the missing scalar constant for the model, arriving at the final result:
\begin{equation}
t_\mathrm{blowup} = 540 \tau_D \phi^{1-d_f/3}\delta^{d_f/3} \epsilon^{3-d_f/3} e^{1/\left(5.6\epsilon\right)}G^{d_f/3-2}.
\label{eq:finalscaling}
\end{equation}
For a colloidal gel \eqref{eq:finalscaling} can be applied to calculate the point in time where the hydrodynamic instability occurs, leading to rapid collapse. 

\subsection{The effect of finite initial channel radius}
\label{sec:seededhole}
During the analysis of the computational model results presented so far, when relating $t_\mathrm{crossover}$ to $t_\mathrm{blowup}$ it was implicitly assumed that $R(0)=0$ and that a streamer is nucleated immediately after the start of the simulation.  However, as will also be discussed in $\S$\ref{sec:future} an initial vertical channel might already be present in the gel network at $t=0$. Or, on observing a gel undergoing collapse, the initial starting point itself may not be well defined in an actual system, especially since in experiments colloidal dispersions do not gel independently from the influence of the gravitational field they are in. Regardless, the model permits a solution for $t_\mathrm{blowup}$ when assuming an initial condition $R_0\neq 0$ for \eqref{eq:ode}. Intuitively, it is expected that catastrophic collapse in a gel with a channel present will occur sooner than in an unperturbed gel. Indeed, using the result in \eqref{eq:full} the model prediction for the shortened finite-time singularity relative to the unperturbed case is found to be:
\begin{equation}
\frac{t_\mathrm{blowup}\lvert_{R_0}}{t_\mathrm{blowup}\lvert_{0}}=\frac{\Gamma\left(2-d_f/3,R_0/R^*\right)}{\Gamma\left(2-d_f/3,0\right)},
\label{eq:holes}
\end{equation}
The ratio of Gamma functions is guaranteed to be less than unity, regardless of the value of $d_f$, in agreement with the intuitive expectation. The only controlling parameter is the value of the initial channel radius $R_0$ \textit{relative} to $R^*$ and this ratio is predicted to be independent of $\phi$. 

\begin{figure}
  \centerline{\includegraphics[width = 0.75\textwidth]{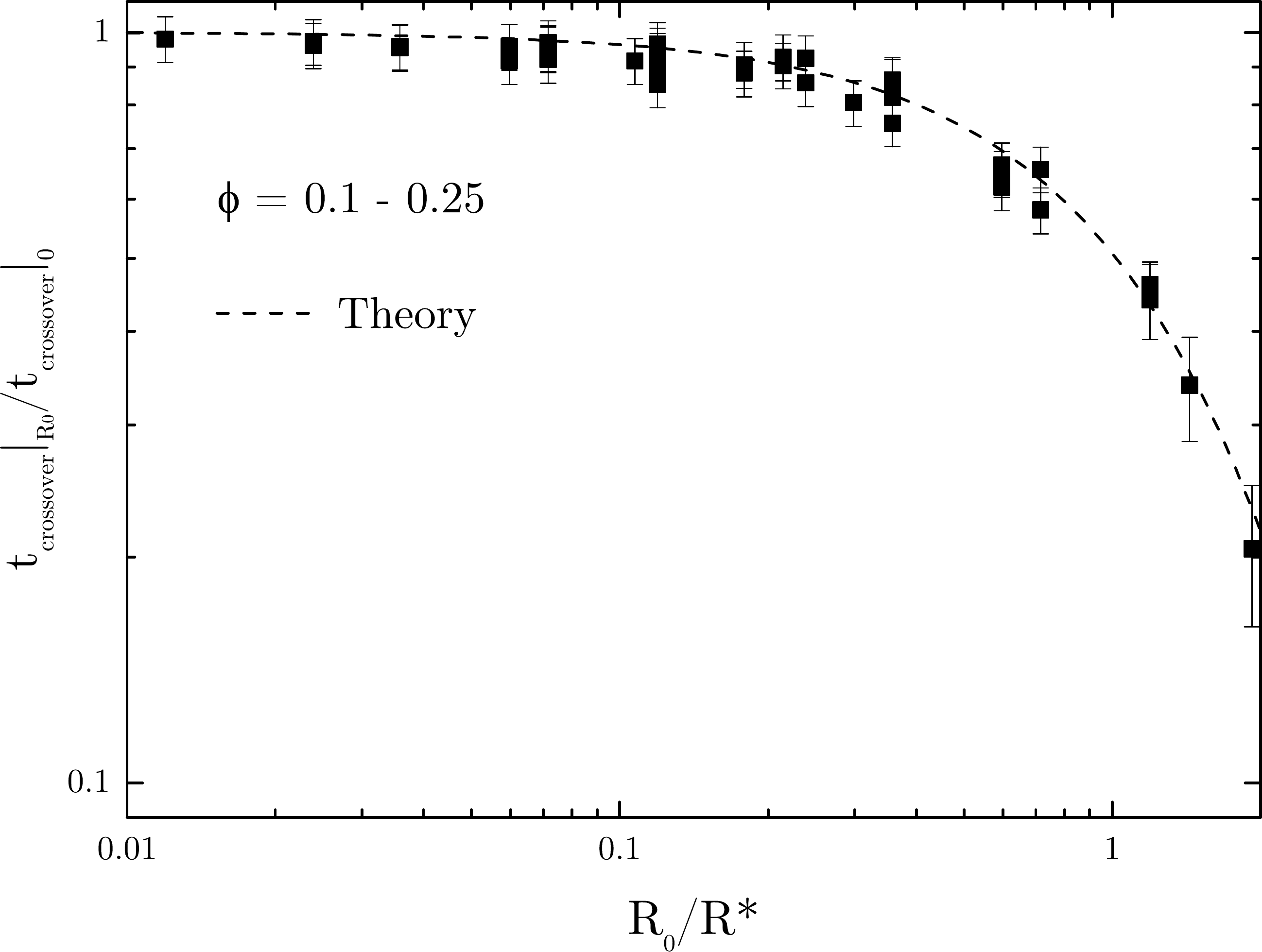}}
  \caption{The measured values of $t_\mathrm{crossover}$ in the presence of a seeded hole at $t=0$ relative to the timescale of an unperturbed sample are plotted as a function of the relative seeded channel radius $R_0/R^*$ for different $\phi$. $R^*$ is set by a combination of $\delta$, $\epsilon$ and $G$. Each data point is the average of three independent simulations with the same unique combination of parameters in the parameter space and error bars represent $95\%$ confidence intervals. The measured values are in good agreement with the model prediction of \eqref{eq:holes} using a value of $d_f=2$(dashed line).\label{fig:hole}}
\end{figure}

To test this prediction we conduct an additional set of simulations of freely settling gels where a vertical cylindrical channel of height $L_{sim}$ and radius $R_0$ devoid of particles is seeded at the centre of the gel at $t=0$. The parameter $R_0/R^*$ is varied systematically. This is achieved both, by increasing $R_0$ relative to the overall size of the gel, and separately, by decreasing $R^*$. Remember, the gravitational length depends on $\delta$, $\epsilon$ and $G$ so that the effect of varying network parameters is directly incorporated. Because of the exponential growth, in order to resolve the decrease in blowup time, we explore a parameter range of 2 orders of magnitude in $R_0/R^*$. As long as the initial pore is small relative to the system size, the assumptions of our model should hold and the onset and dynamics of collapse should only be determined by the radius of the streamer. In each simulation the crossover time with a seeded pore is measured and compared to the corresponding crossover time in an unseeded gel at the conditions. We plot their ratio in figure \ref{fig:hole} along with the prediction of \eqref{eq:holes}.

We find an excellent agreement between \eqref{eq:holes} and the simulation results. For small seeded channel sizes, $ R(0) \ll R^* $, the collapse times are nearly the same with only a negligible decrease, as stated in $\S$\ref{sec:finite}. In contrast, for $R_0/ R^*~O(1)$, $t_\mathrm{crossover}\lvert_{R_0}$ falls off due to the activated hopping of particles off the network, as expected from the theory. Simulations show that the seeded channel provides a free path for fluid back flow when the gel is settling. Particles on the network-streamer interface are immediately ripped off the network and swept upwards.  This provides further evidence that the growth of a streamer in a settling gel is indeed the cause for the rapid instability leading to collapse. As we have shown, the predictions of the simple model of streamer growth agrees well with the collapse dynamics observed in simulations for a large range of parameter values and initial gel states.  

\subsection{Comparison with experimental observations}
\label{sec:experiment}
So far it was shown that the model correctly captures the mechanism of collapse seen in dynamic simulations. Here we show that our theory also describes the gel collapse behaviour in two vastly different experimental systems by fitting available data of the evolution of the observed streamer radius as a function of time to our model. 

\citet{starrs2002collapse} studied a system of polydispserse PMMA particles that were induced to gel in the presence of polystyrene depletants in a tetralin and cis-decalin solvent blend. In so-called "weak" gels streamers formed, and the gel exhibited catastrophic collapse following a hydrodynamic instability. The weak gels were differentiated from "strong" gels considered in the study by the lack of hydrodynamic instability and the observation of steady, poro-elastic compression rather than a dramatic collapse. \citet{secchi2014time} looked at an aqueous dispersions of spherical MFA particles, a copolymer of tetrafluoroethylene and perfluoromethylvinylether. Depletion interactions were induced by the addition of a surfactant that forms globular micelles leading to a short-ranged attraction. Here, the onset and subsequent radial growth of streamers destabilizing the network was also observed. In both cases, snapshots depicting the increasing size of the streamer with time were included by the authors, providing two experimental data sets for the evolution of the streamer radius, $R(t)$.  Table \ref{tab:parameters} provides the parameters for the two experimental systems.

\begin{table}
  \begin{center}
\def~{\hphantom{0}}
  \begin{tabular}{l|c|c}
   Parameter    & Starrs \etal     &  Secchi \etal\\[3pt]\hline
       Attraction width $\Delta$ (nm) & 17 & 4\\
       Attraction strength $U$ ($kT$)   & 6.5 & 12\\
       Density mismatch $\Delta \rho$ (g/cm$^{3}$)  & 0.26 & 1.14\\
       Particle radius $a$ (nm)   & 196 & 90\\
       Solvent viscosity $\eta$ (mPa s)   & 2.5 & 1.0\\
       Temperature $T$ ($^o$C)   & 25 & 23\\
       Volume fraction $\phi$ (-)   & 0.2 & 0.04\\

  \end{tabular}
  \caption{Experimental parameters of collapsing gels as found in \citet{starrs2002collapse} and \citet{secchi2014time}.}
  \label{tab:parameters}
  \end{center}
\end{table}

\begin{figure}
\begin{subfigure}[b]{0.45\textwidth}
               \includegraphics[width=\linewidth]{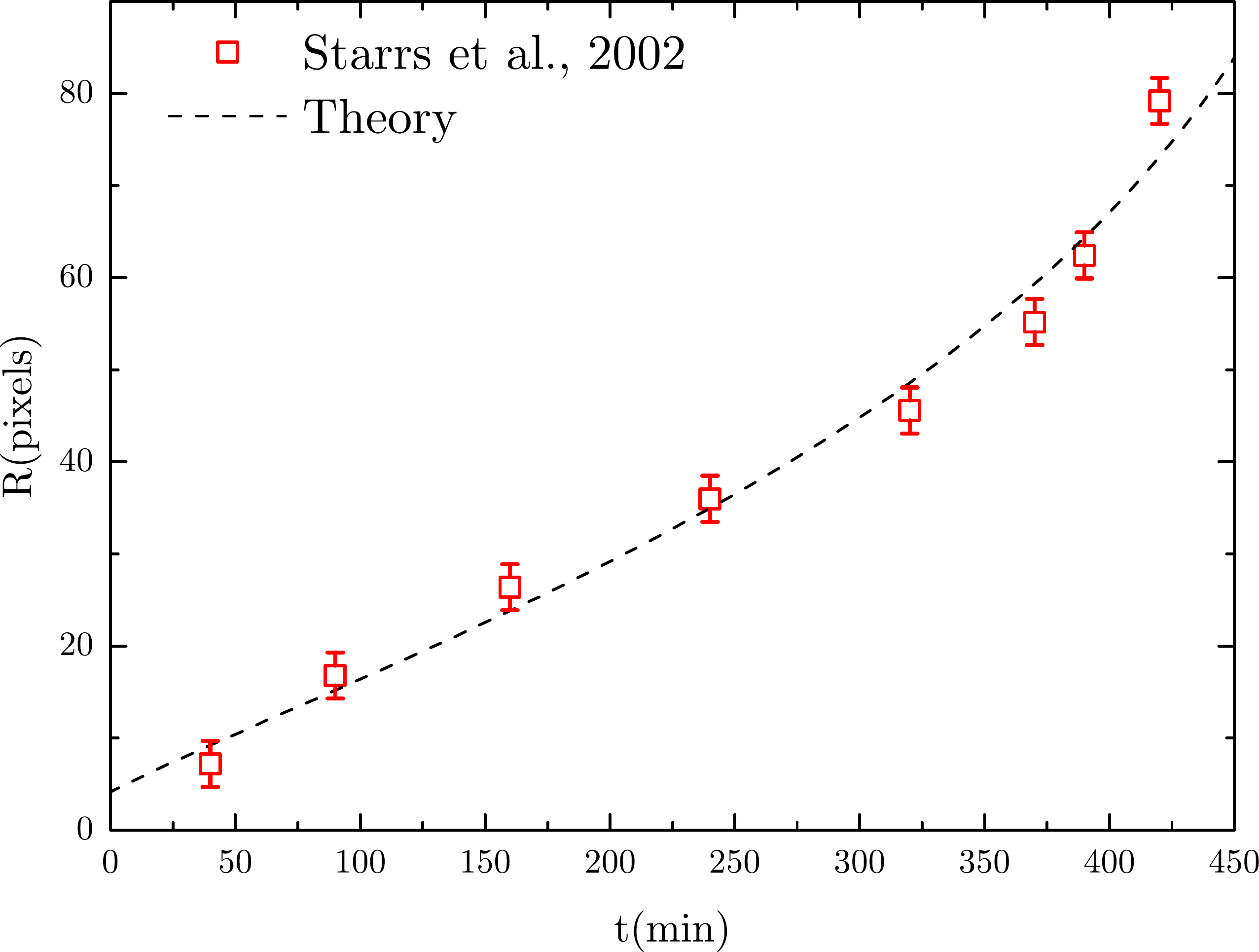}
               \caption{Comparison of data published by \citet{starrs2002collapse} (open squares) and the best-fit prediction of the model (dashed line).\label{fig:Starrs} }	    
				\end{subfigure}\hfill
				\begin{subfigure}[b]{0.45\textwidth}
               \includegraphics[width=\linewidth]{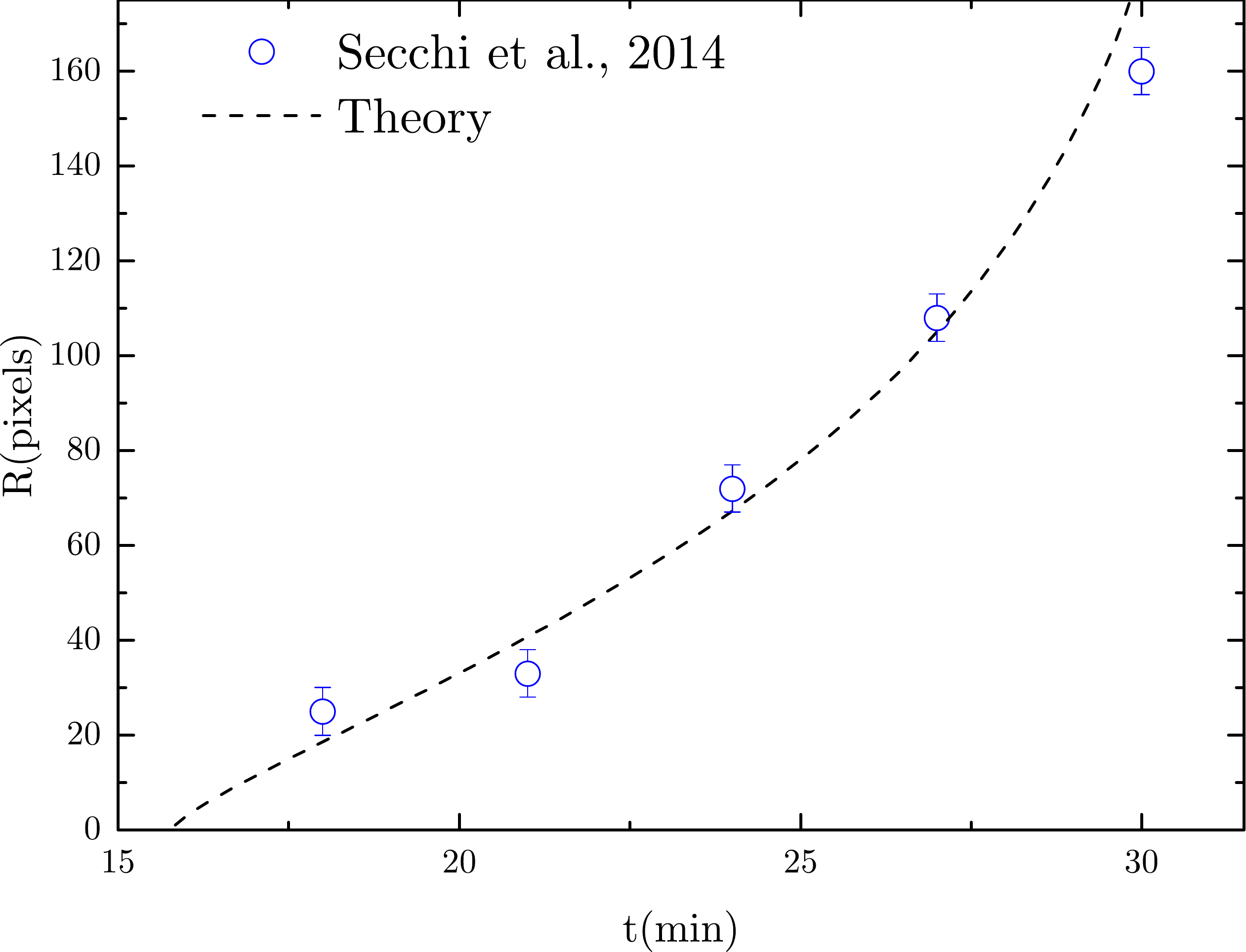}
  				\caption{Comparison of data published by \citet{secchi2014time} (open circles) and the best-fit prediction of the model (dashed line).\label{fig:Secchi}}		
        		\end{subfigure}%
        \caption{Time evolution of streamer radius $R$ as extracted from experimental data published in the literature and the best-fit predictions of our model.\label{fig:expboth}}
\end{figure}

Assuming an initial pore size, $ R(0) = 0 $, equation \eqref{eq:ode} has solution:
\begin{equation}
\frac{t_\mathrm{blowup} - t}{ t_\mathrm{blowup}} = \frac{\Gamma( 2 - d_f / 3, R(t)/R^* )}{\Gamma( 2 - d_f / 3, 0 )}, \label{eq:dynamics}
\end{equation}
which has three parameters: $ t_\mathrm{blowup} $, $ R^* $, and $ d_f $.  As seen earlier, the model predictions are not sensitive to the value of $d_f$ in the range of $1.7-2.3$ and so in the absence of any further information from the experiments, we assume that $d_f=2$ for both networks, a typical value for dilute colloidal gels\citep{zaccarelli2007colloidal}. Therefore, it remains to obtain a best-fit of \eqref{eq:dynamics} to the two experimental data sets to extract the parameters $t_\mathrm{blowup}$ and $R^*$. Figure \ref{fig:expboth} compares the experimentally observed streamer radius to the best-fit model predictions.  The measured dynamics for both experimental systems follow the model quite well, and table \ref{tab:exp} provides the best fit values for $ t_\mathrm{blowup} $ and $ R^* $.

The quantities $ t_\mathrm{blowup} $ and $ R^* $ can be estimated from experimental parameters independent of the model fit. Using the values of the experimental quantities in table \ref{tab:parameters}, $t_\mathrm{blowup}$ and $ R^* $ are computed directly using \eqref{eq:rstar} and \eqref{eq:finalscaling} assuming again $d_f = 2$ and taking the coefficient $ d_4 = 1 $. These estimates are reported in table \ref{tab:exp}. In the case of the work by \citet{starrs2002collapse} the best-fit values for both parameters are in good agreement with what can be computed \textit{a priori}. This would suggest that the proposed model subsumes all essential factors contributing to streamer growth and can accurately describe the dynamics of network erosion and collapse. In contrast, the estimates of $R^*$ and $t_\mathrm{blowup}$ for the other example are an order of magnitude larger than the corresponding best-fit values.  $ R^* $ exceeds the dimensions of the gel and $ t_\mathrm{blowup} $ exceeds the observation window reported\citep{secchi2014time}. Since the dynamics are well captured by the best-fit, this would indicate that there is no seeded pore in the gel.  Instead, we conclude that there is some uncertainty in the experimental parameters. While we have assumed that the characteristic building blocks of the gel in the experiments are individual MFA particles, \citet{secchi2014time} note that there is strong evidence to suggest that the particles move in aggregated clusters. As the model displays a high sensitivity to the size of particles within the gel (see $\S$\ref{sec:twotimes}) the value of the hydrodynamic radius $a$ can significantly impact the quality of the predicted blow-up time.  Indeed, using $a=3\times90$nm as the characteristics size, we recover estimates of $R^*$ and $t_\mathrm{blowup}$, shown in the last row of table \ref{tab:exp} that are both physically reasonable and agree with the best-fit values.

\begin{table}
  \begin{center}
\def~{\hphantom{0}}
  \begin{tabular}{lcc|cc}
        & \multicolumn{2}{c|}{Starrs \etal}     &  \multicolumn{2}{c}{Secchi \etal}\\[3pt]
        & $t_\mathrm{blowup}$(min) & $R^*$(mm) & $t_\mathrm{blowup}$(min) & $R^*(mm)$ \\ [3pt] \hline
       Best-fit   & 593$\pm$11 & 4.85$\pm$0.1 & 34$\pm$2 & 2.02$\pm$0.15\\
       Estimate from Parameters   & 623 & 4.94 & 986 & 22.5\\
       & & & 37 & 2.51
  \end{tabular}
  \caption{Values of $t_\mathrm{blowup}$ and $R^*$ as found from the model best-fit to the experimental data and their estimates based on details of the experimental parameters in table \ref{tab:parameters}. In the case of \citet{secchi2014time} estimates are based on the primary particle radius and three time the size for $a$.}
  \label{tab:exp}
  \end{center}
\end{table}

\begin{figure}
  \centerline{\includegraphics[width = 0.75\textwidth]{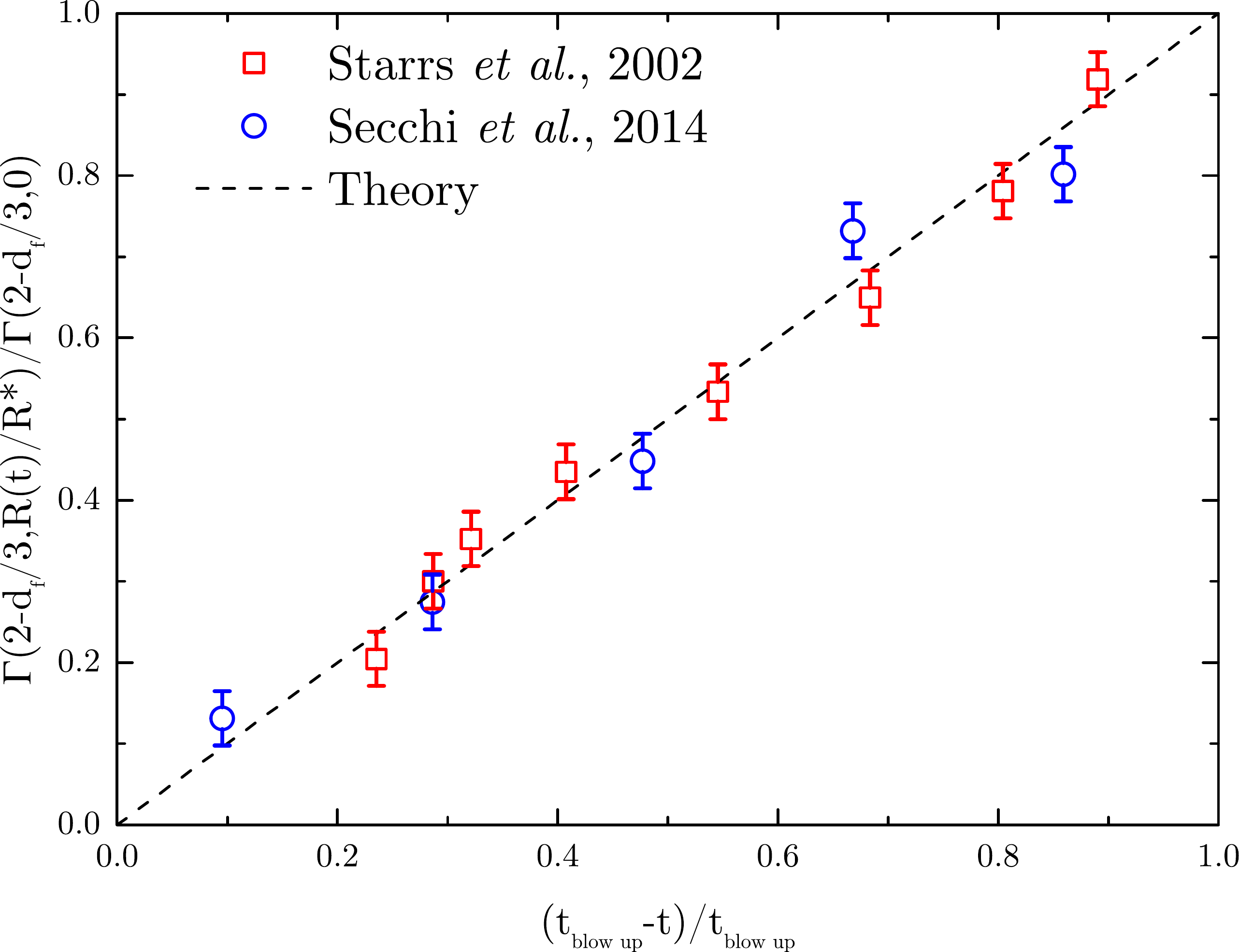}}
  \caption{Parity plot of the predicted and the measured distance to blowup (see text for details) for the two very different experimental systems.  The theory is able to capture the underlying collapse dynamics well, independent of the experimental details. \label{fig:exp_all}}
\end{figure}

As a result of the exponential growth rate of the streamer radius, the value of $R(t)$ and the dynamics in the vicinity of blowup are very sensitive to $t$ and change rapidly, as seen in figure \ref{fig:expboth}. While $t_\mathrm{blowup} $ and $R^*$ depend on the experimental parameters and are unique to each study, \eqref{eq:dynamics} suggests that $ \Gamma( 2 - d_f/3, R(t)/R^*) $ will exhibit a universal linear dependence on the distance to blowup, $\left(t_\mathrm{blowup}-t\right) $. Figure \ref{fig:exp_all} plots the experimentally measured streamer radii as a function of distance to blowup, and we observe indeed a one-to-one parity as expected from theory. This indicates that the model for streamer growth contains the necessary dynamics to understand and track the hydrodynamic instability leading to the collapse of the network in both colloidal gels.

\section{Discussion}
\label{sec:discussion}
Gravitational collapse is an intricate process during which the microstructure of the gel undergoes numerous changes. These culminate in the break down of the hierarchical network structure and macroscopic structural failure. A number of experiments observe a "sudden" collapse in which a slowly and steadily moving interface suddenly accelerates and falls. The model we have presented describes in detail a process through which erosion of the network and the growth of streamers in a freely settling gel can lead to such a perceived sudden change in material properties.  We have shown that these significant rearrangements in the microstructure do not manifest themselves on a macroscopic level until a time point very close to a singularity in the streamer growth. From the point of view of applying this model to real gels, a good theory of gravitational collapse has to be able to explain and predict the ultimate parameter of interest, $t_\mathrm{blowup}$ as presented here. The model we have proposed relates $t_\mathrm{blowup}$ to the engineerable network and solvent properties. Both dynamic simulations and comparisons with previously published experiments have confirmed that the model reproduces quantitatively the collapse dynamics of these gels.

\subsection{Stability is dictated by a competition between two timescales}
\label{sec:twotimes}
The time scale $t_\mathrm{blowup}$ is an intrinsic property of each freely settling gel network and characterizes how long the gel remains stable before hydrodynamic back flows erode network integrity.  The proposed model suggests that this hydrodynamic instability will occur at a definite point in time, after the beginning of free settling. According to the model, no gel is immune to this instability -- yet stable gels can be engineered. Certain strong gels remain stable against gravitational stresses for years, while only compacting mildly\citep{poon2002physics, manley2005gravitational, teece2014gels}. As described earlier, these strong gels exhibit a slow uniform compression under gravity. By compacting steadily in time, the gel becomes denser and strong, allowing it to fulfill its engineered purpose. The slow condensation process is well described by the theory of poro-elasticity, in which forces responsible for mechanical compression balance with the drag due to \emph{uniform} fluid back flow and the elastic stresses within the compacting network. Poro-elastic settling occurs over a characteristic timescale\citep{manley2005gravitational}:
\begin{equation}
\label{eq:te}
t_\mathrm{poro-elastic}\sim \frac{\eta h_0^2}{\kappa E}\sim\frac{\eta h_0^2 \phi^{2/(3-d_f)}}{a^2 E},
\end{equation}
which measures the time required for the compressing network to develop sufficient strength to support its own weight. Here $h_0$ is the initial height of the gel, $E$ is its elastic modulus and the permeability $\kappa\sim a^2/\phi^{2/(3-d_f)} $ with an $O(1)$ prefactor, will depend on the porosity of the network.  

All networks in which a density mismatch between fluid solvent and solid particles is present, will initially sediment in this manner. However, in weak gels the process of  poro-elastic compression is interrupted by the formation of streamers which leads to subsequent rapid settling. In this framework then what distinguishes strong from weak gels under gravity is the ratio of two timescales:
\begin{equation}
T=\frac{t_\mathrm{blowup}}{t_\mathrm{poro-elastic}},
\end{equation}
which provides a criterion for deciding whether a network will exhibit poro-elastic compression or streamer mediated collapse. When $T\gg 1$ a colloidal network will exhibit characteristics of poro-elastic compression.  Initially, much of the network will be in a mode of free settling which can produce streamers, but the time scale for streamer formation, $ t_\mathrm{blowup} $ is too long for such pores to form.  Instead, the settling will come to an end when the compressed gel has developed enough strength to support itself.  In contrast, for $T\ll 1$ the gel will settle, but does not densify quickly enough.  Instead, streamers nucleate within the gel, eliminate any elastic resistance through erosion of the network, and result in rapid collapse. 

This criterion is also supported by experimental observations.  In addition to the set of experiments on weak gels, \citet{starrs2002collapse} also studied the collapse of gels with stronger inter-particle attractions. These gels did not collapse, but instead underwent steady, poro-elastic compression, which arrested in a more compact, and stable state. The timescale for this consolidation process here was reported to be $t_\mathrm{poro-elastic}= 40$ hours. For the set of experimental parameters corresponding to this strong gel, our model predicts $t_\mathrm{blowup}\approx 44$ hours so that $T>1$.  From this ratio of time scales, we would expect that the gel should remain intact. In contrast, consider the value of $ T $ anticipated for the weak, collapsing gels that were studied\citep{starrs2002collapse}.  The completion of poro-elastic compression was not be observed in the experiments, but we estimate $t_\mathrm{poro-elastic}\approx 32 $ hours using the assumption that the network elastic modulus decreases in proportion with the strength of the interparticle attraction. From fitting to the model and independent calculation, we determined that $t_\mathrm{blowup}\approx 10$ hours.  Consequently $T<1$, and the gel is expected to be unstable.


Equations \eqref{eq:scaling} and \eqref{eq:te} show that $ t_\mathrm{blowup} $ and $ t_\mathrm{poroelastic} $ are only functions of material properties. Therefore it should be possible to evaluate $T$ in advance and predict the stability of a proposed experimental system without any detailed experimentation.  Especially useful is understanding how specific parameters influence this ratio.  For instance, consider the dependence on particle size.  We have shown that $t_\mathrm{blowup}\sim a^{d_f-5}$.  Because the elastic modulus of the network depends on its mesh size, we conclude that $E\sim a^{-3} U/kT$, and the poro-elastic timescale is linearly proportional with the particle radius: $t_\mathrm{poro-elastic}\sim a$. Therefore the ratio of timescales depends on $ a $ as: $T\sim a^{d_f-6}$, suggesting that the stability of colloidal gels is strongly controlled by the primary particle size.  In fact, decreasing the particle size will drive the network toward pure poro-elastic compression.  This ratio also helps to explain why the stability of a gel is so sensitive to changes in $U$\citep{starrs2002collapse}. From \eqref{eq:finalscaling}, the streamer formation time depends most dominantly exponentially on $ U $, while from \eqref{eq:te}, the poro-elastic timescale scales with the inverse of $ U $.  Consequently, $T \sim e^{U/(5.6kT)}$, and small changes to the strength of attraction in $U$ will lead to large changes in the ratio of timescales that significantly alter the stability of the gel. Finally, the hydrodynamic instability is dominated by the activated erosion process driving the streamer growth and is thus sensitive only to intrinsic properties of the network.  In contrast, the poro-elastic compression timescale depends on the initial height of the gel. Thus, $T\sim h_0^{-2}$, so that taller gels are more susceptible to the hydrodynamic instability.  In laboratory experiments, this is known and shorter gels, which may appear stronger, are often avoided when studying the collapse phenomena\citep{allain1995aggregation, starrs2002collapse, kilfoil2003dynamics, manley2005gravitational}.

\begin{figure}
  \centerline{\includegraphics[width = 0.75\textwidth]{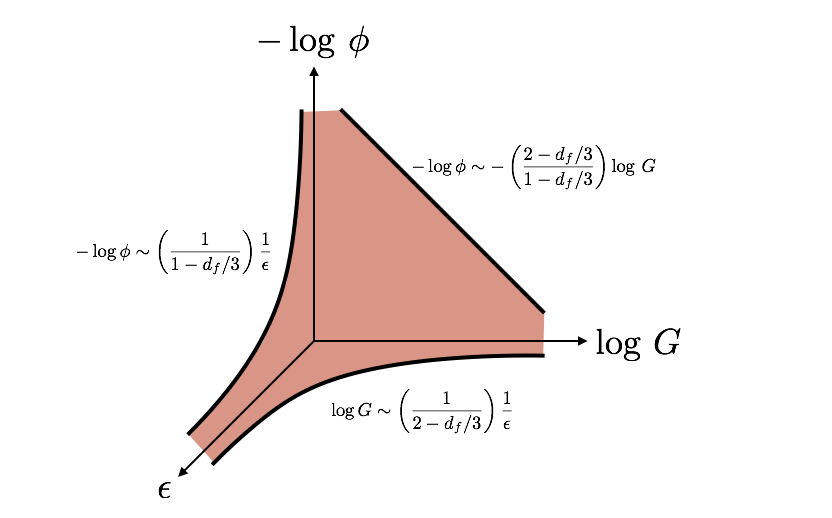}}
  \caption{The stability state diagram marks the continuous region of stability(shaded) and distinguishes it from catastrophic instability beyond (here only a three dimensional cross section of the entire parameter space is shown). In the stable region $t_\mathrm{blowup}$ is larger than the desired $t_\mathrm{shelf life}$ of a given product.\label{fig:stability}}
\end{figure}

In many practical applications, whether to avoid the collapse of yoghurt in a cup or the failure of a gel proppant in a fracking channel, the required shelf life of the material, $t_\mathrm{shelf life}$, is a well defined finite quantity. In terms of the model discussed here, the requirement for stability necessitates that the blowup time exceeds the application or user defined shelf life. Thus, even if it is not possible to choose parameters such that $T\ll 1$ and achieve indefinite stability, at least material properties can be tuned with the goal of an extended lifetime so that $t_\mathrm{shelf life}\leq t_\mathrm{blowup}$, which defines the desired minimum value for the blowup time. This way it can be ensured that the hydrodynamic instability will only set in once the gel network has fulfilled its role past its required lifetime. In practice then, given a use case defined constant $min\left(t_\mathrm{blowup}\right)=t_\mathrm{shelf life}$, \eqref{eq:finalscaling} determines how the values of the network properties must be chosen. In figure \ref{fig:stability} we present a three dimensional subset of the four dimensional parameter space of $\delta$, $\epsilon$, $G$ and $\phi$, which represents the trade-offs that have to be considered in material engineering of stable gel networks. In the continuous stable region characterized by high volume fractions, low gravitational Mason numbers and high network strengths, the blowup time exceeds the desired shelf life of the specific product under consideration. The boundaries between the stable and unstable region will be highly non-linear, but are defined by the respective parameter pairs and the model predictions.  Such a stability diagram enables rational selection of materials or engineering of colloidal interactions, both of which involve trade-offs in the space of network properties.

Consider yet another application, where colloidal gel networks are used in processes of sludge reduction and dewatering. In such contaminated site remediation programs, the material essentially acts as a filter used to halt the flow of the pollutants while the whole network is under fluid driven compression\citep{northcott2005dewatering}. Here, the effective gravitational Mason number, $G$, is controlled by the process operator through the choice of $|\nabla p|$.  This parameter choice ultimately determines the time point at which the network becomes unstable.  In this application, the other relevant timescale is the duration of the dewatering process\citep{northcott2005contaminated}, $t_\mathrm{process}$, controlling the total amount of sludge that is processed. The maximum rate at which the water can be treated safely without any network failure, will be set by the ratio of these two timescales, $ t_\mathrm{blowup}/t_\mathrm{process}$ and the same stability criteria apply. The proposed model may aid in selecting the correct operating conditions for such remediation activities.

\subsection{Model improvements and future work}
\label{sec:future}
As we have shown, the model adequately predicts the observed dynamics in simulations and explains the onset of sudden collapse in experiments. However, in developing the model a few simplifying assumptions had to be made: chiefly about the number of particles within the streamers that form, the process of activated bond breaking, the fractal nature of the gel, and the boundary conditions imposed on the network. Here we revisit some of these assumptions and discuss future work to resolve remaining issues.

During the growth of the streamer, as particles are broken off the fractal network at the channel interface and swept upwards due to back flow, local gradients in particle concentration will be established and the distribution of particles within the streamer should not be uniform as assumed when constructing \eqref{eq:drfirst}. However, as discussed during the model development, for a sufficiently large control volume around the streamer and the porous gel network, the number density of particles in the interior of the streamer due to mass conservation will have to be equal to its bulk value. In this context, a sufficiently large control volume means a streamer whose height is very large compared to the typical mesh size of the gel network. On that length scale, the expression used for the number of particles on the surface of the network will be valid, and mass conservation will ensure that the model predicts a consistent number of particles entering the streamer per unit time.

The growth of the streamer is driven by the viscous drag exerted by the fluid back flow resulting in activated bond breaking of individual inter-particle bonds, which drives the growth in $R$. The bond breaking rate, which scales as: $e^{-\left(1/\epsilon\right)^{1/d_3}}$, reflects the particle escape probability from the potential well. We assumed that the effective activation energy setting this rate will not be the strength $U$ of a single inter-particle bond, but an undefined scalar multiple. A value of $d_3>1$, which was obtained from the simulation results in figure \ref{fig:eps}, would suggest that the bonds are weaker than in the case of an escape from a single pairwise bond. In fact, when a bond in the gel is broken, a single particle or a cluster of particles may detach from the network and enter the streamer. As a result, it is probably more useful to think of the activated escape rate as being set by several independent bond breaking events, any one of which might free some portion of a cluster to follow the back flow.  Thus the $ 1 / d_3 $ power in the proposed Kramers hopping expression allows for more attempts at bond breakage per unit time.  To more precisely account for this effect, we would need to track the statistics of clusters entering the streamer in order to weight the flux by the appropriate number of particles detaching from the gel.  This is planned for future work. 


The five dimensionless parameters in table \ref{tab:groups} fully characterize the model gel. Of these, the fractal dimension $d_f$ is the most difficult to control as it is a characteristic of the kinetically arrested percolated network and how that network was formed.  In principle this could be predicted, but in practice it is more likely to be measured once the gel has formed. In experiments, $d_f$ is obtained from the power-law growth of the structure factor at low wavelengths, whereas in simulations due to the finite system size, as is the case here, the box-counting method is employed as a surrogate. Thus, $d_f$ is not known in advance and \eqref{eq:finalscaling} cannot be predictive in the strictest sense. For the model to be useful, a value of $d_f$ has to be assumed, which can be justified by prior experience with the specific gel under study and the fact that for random percolated gels the fractal dimension is typically in the range $d_f\approx 1.7-2.3$\citep{zaccarelli2007colloidal}. In fact, when fitting the model to the experimental streamer growth data, in the absence of any measurements, the assumption of $d_f=2$ resulted in good agreement. Clearly, an accurate knowledge of $d_f$ could improve this and result in a better match between parameter estimates and the model predictions. We were able to validate the scaling of the blowup time by studying $t_\mathrm{crossover}$ as a function of $\delta$, $\epsilon$, $G$ and $\phi$ in figure \ref{fig:details}.  However, \eqref{eq:finalscaling} is not especially sensitive to the value of $d_f$ due to the limed range of variability, and we could not control $ d_f $ explicitly.  Therefore, dependence of the model on the value of $ d_f $ remains to be confirmed.

Since the model as presented here considers a freely settling gel, it neglects the processes occurring at both the top and bottom interfaces in contact with the supernatant and compacting region, respectively\citep{padmanabhan2018gravitational}.  As shown by \citet{buscall1987consolidation} and repeatedly observed in experiments, the majority of the collapsing gel is in a mode of free fall and macroscopically unchanging before collapse\citep{secchi2014time}. Since the particles at the bottom of the gel are unable to support the weight of the network above it, a concentrated foot grows at the bottom of the container by continuous, slow compaction. However, this process is part of the poro-elastic compression, and it is largely agreed that the compaction occurs independent of the events leading to the hydrodynamic instability in the freely settling region. 

In contrast, it has been suggested that the origins of delayed collapse are related to the free surface at the top of the gel. Interactions between the meniscus and the container walls may delay the settling as the network is pinned to the top interface. The surface tension holding the colloidal particles at the air-solvent interface can be significant with energy per unit surface area on the order of: $ 10^3-10^6 kT/a^2$\citep{binks2006colloidal}. In contrast, particles in the layers beneath the gel surface are held in place solely by inter-particle bonds that cannot support the tension due to the weight of the gel network hanging below. Thus, while the forces on the particle contact line can suspend a layer of particles at the interface\citep{secchi2014time}, the entire network is not pinned and can detach readily without producing a detectable delay preceding collapse. 


However, it may be the case that the interface with the supernatant has a role to play in seeding an initial streamer through the network. Colloidal particles at the air-solvent interface are in constant motion owing to thermal motion\citep{boniello2015brownian}. Recently it has been observed that particles coalesce and form concentrated clusters at the top interface\citep{harich2016gravitational}. These fragments compact, break off and can fall through the network to create the channels whose growth our model could describe. Indeed, the model presented here lacks an exact description of how an initial channel is seeded. Gels are kinetically arrested materials with structural heterogeneities.  Therefore, it was assumed that streamers form from natural density fluctuations. It is entirely plausible and fully compatible with the model that paths of preferred fluid back flow are seeded by other restructuring and aggregation processes, for example: bubbles which rise through the network, or foreign objects and debris falling through the gel\citep{senis2001settling, teece2014gels, harich2016gravitational}.  As shown in $\S$\ref{sec:seededhole} with controlled simulations, a seeded initial channel radius $R_0$ produces the same collapse dynamics, which are universally described by the phenomenological model.  Furthermore, consider the network stability in a range of industrial applications, where very often, the choice of formulation or processing history of the gel will result in a number of holes and channels that extend across the sample and may be distributed randomly. In such a case the overall stability of the network will be set by the growth rate of the streamer expanding the fastest. Thus the onset of the hydrodynamic instability can only be determined with knowledge of the heterogenous state of the gel. However, as long as the distribution of initial channel sizes can be estimated reasonably accurately, a distribution of blowup times can be inferred and a survival probability used to characterize the sample lifetime.

Gravitational collapse of gels is a complex phenomenon with many dynamic processes occurring throughout the network simultaneously. The micromechanical model described here sheds light on one central aspect, the rapid growth of streamers leading to the instability and collapse of freely settling gels. Nonetheless, new approaches are required to be able to study the settling processes at the gel boundaries and free interfaces. In the freely settling mode hydrostatic equilibrium cannot be established, so the eventual arrest of the settling process will be intimately coupled to the interactions with the container walls. While dynamic simulations provide a useful tool to interrogate the microstructure, it is the interplay of inter-particle interactions and hydrodynamics that leads to intricate settling scenarios.  Therefore and importantly, hydrodynamic interactions have to be accurately incorporated into computer models in order to describe the basic phenomenology. Modelling hydrodynamic interactions in colloidal dispersions near interfaces and walls is a challenging task. The development of computational tools for fast simulations at sufficient scale to tackle detailed interactions among the particles and with container boundaries is an active area of research\citep{fiore2018rapid, fiore2018stokes}. Future investigations should seek to conduct new experiments and computer modelling of the same colloidal gels in order to enable more careful comparisons of the detailed dynamics between model predictions, experiments and simulations. 

\section{Conclusions}
\label{sec:conclusions}
The catastrophic collapse of colloidal gels settling under their own weight remains a major engineering challenge in many areas of industry and science, from personal care and foodstuffs through industrial proppants to biomedical applications. We have shown with models and simulations that the sudden transition from slow uniform settling to rapid and complete loss of network integrity has a unified origin in these applications. Over the last decades, careful experiments have advanced an understanding of the restructuring preceding collapse: bond rearrangements and breakage lead to the formation of open streamers through the network. Our simulations have enabled the direct observation of fluid back flow through these streamers and the effects of the viscous drag that the back flow exerts on the gel network.  These stresses erode the network and lead to a hydrodynamic instability that terminates in failure of the gel.

We developed a new phenomenological model for the evolution of streamers embedded in a freely settling colloidal network. The model describes the process of streamer growth due to fluid back flow, which strips particle from walls of the streamer.  The rate of erosion increases exponentially with the streamer radius so that the model exhibits a finite-time blowup: At a finite point in time, the radius of the streamer is infinite.  We correlate this point in time with the onset of catastrophic failure in the gel. This timescale is related directly to dimensionless groups describing the network: the ratio of buoyant forces to network strength, the particle volume fraction, the strength of inter-particle bonds relative to the thermal forces acting on the particles, and the relative range of the pairwise attraction. 

Extensive Brownian dynamics simulations of hydrodynamically interacting, freely settling, attractive colloidal gel networks show that the rapid increase of the streamer volume in the gel coincides with increased settling velocities during collapse. The time for onset of accelerated settling scales with network parameters as predicted by the model, and we have demonstrated a direct parity between the model blowup time and this critical time point in simulations. The extensive parameter sweep conducted in simulations is used to determine the unknown constant of proportionality of the phenomenological model, which is necessary to make quantitative predictions. The predicted evolution of streamer radius with waiting time is also shown to successfully capture the collapse dynamics for two different published experimental systems.

The model considers a gel in a mode of free settling and neglects the effects of container walls and the processes occurring at both the top and bottom interfaces in contact with the supernatant and compacting cake region, respectively. Regardless, the model accurately predicts the dynamics of the hydrodynamic instability leading to collapse in freely settling gels. We find that the critical feature demarcating strong from weak gels under collapse is the ratio of the poro-elastic compression timescale to the finite-time blowup. Since both processes are intrinsic to any gel under gravitational load, strong gels are the ones where poro-elastic compression proceeds to completion before the onset of the hydrodynamic instability. Therefore the key to achieving longer shelf lives is to engineer and tune network properties until the blowup time exceeds the poro-elastic time scale, the user defined shelf life of the product, or the relevant process time of the application. With the concepts presented in this work and the newly developed model, stability of colloidal networks can be rationally engineered. 
\linebreak

The authors acknowledge helpful conversations with Professors Gareth McKinley, Wilson Poon and Jan Vermant, and funding provided by the ACS Petroleum Research Fund (grant no. 56719-DNI9) and the Institute for Soldier Nanotechnology at MIT.

\bibliographystyle{jfm}

\begin{thebibliography}{57}
\expandafter\ifx\csname natexlab\endcsname\relax\def\natexlab#1{#1}\fi
\def\au#1{#1} \def\ed#1{#1} \def\yr#1{#1}\def\at#1{#1}\def\jt#1{\textit{#1}}
  \def\bt#1{#1}\def\bvol#1{\textbf{#1}} \def\vol#1{#1} \def\pg#1{#1}
  \def\publ#1{#1}\def\arxiv#1{#1}\def\org#1{#1}\def\st#1{\textit{#1}}

\bibitem[Acrivos \& Goddard(1965)]{acrivos1965asymptotic}
{\sc \au{Acrivos, Andreas} \& \au{Goddard, JD}} \yr{1965}  \at{Asymptotic
  expansions for laminar forced-convection heat and mass transfer}.
  \jt{Journal of Fluid Mechanics}  \bvol{23}~(2),  \pg{273--291}.

\bibitem[Allain {\em et~al.\/}(1995)Allain, Cloitre \&
  Wafra]{allain1995aggregation}
{\sc \au{Allain, C}, \au{Cloitre, M} \& \au{Wafra, M}} \yr{1995}
  \at{Aggregation and sedimentation in colloidal suspensions}.  \jt{Physical
  review letters}  \bvol{74}~(8),  \pg{1478}.

\bibitem[Asakura \& Oosawa(1958)]{Asakura1958}
{\sc \au{Asakura, Sho} \& \au{Oosawa, Fumio}} \yr{1958}  \at{Interaction
  between particles suspended in solutions of macromolecules}.  \jt{Journal of
  Polymer Science}  \bvol{33}~(126),  \pg{183--192}.

\bibitem[Bai {\em et~al.\/}(2007)Bai, Liu, Coste, Li {\em
  et~al.\/}]{bai2007preformed}
{\sc \au{Bai, Baojun}, \au{Liu, Yuzhang}, \au{Coste, Jean-Paul}, \au{Li,
  Liangxiong} \& \au{others}} \yr{2007}  \at{Preformed particle gel for
  conformance control: transport mechanism through porous media}.  \jt{SPE
  Reservoir Evaluation \& Engineering}  \bvol{10}~(02),  \pg{176--184}.

\bibitem[Bailey {\em et~al.\/}(2007)Bailey, Poon, Christianson, Schofield,
  Gasser, Prasad, Manley, Segre, Cipelletti, Meyer {\em
  et~al.\/}]{bailey2007spinodal}
{\sc \au{Bailey, AE}, \au{Poon, WCK}, \au{Christianson, Rebecca~J},
  \au{Schofield, AB}, \au{Gasser, U}, \au{Prasad, V}, \au{Manley, Suliana},
  \au{Segre, PN}, \au{Cipelletti, Luca}, \au{Meyer, WV} \& \au{others}}
  \yr{2007}  \at{Spinodal decomposition in a model colloid-polymer mixture in
  microgravity}.  \jt{Physical review letters}  \bvol{99}~(20),  \pg{205701}.

\bibitem[Bartlett {\em et~al.\/}(2012)Bartlett, Teece \&
  Faers]{bartlett2012sudden}
{\sc \au{Bartlett, Paul}, \au{Teece, Lisa~J} \& \au{Faers, Malcolm~A}}
  \yr{2012}  \at{Sudden collapse of a colloidal gel}.  \jt{Physical Review E}
  \bvol{85}~(2),  \pg{021404}.

\bibitem[Beavers \& Joseph(1967)]{beavers1967boundary}
{\sc \au{Beavers, Gordon~S} \& \au{Joseph, Daniel~D}} \yr{1967}  \at{Boundary
  conditions at a naturally permeable wall}.  \jt{Journal of fluid mechanics}
  \bvol{30}~(1),  \pg{197--207}.

\bibitem[Binks \& Horozov(2006)]{binks2006colloidal}
{\sc \au{Binks, Bernard~P} \& \au{Horozov, Tommy~S}} \yr{2006} {\em Colloidal
  particles at liquid interfaces\/}.  \publ{Cambridge University Press}.

\bibitem[Blijdenstein {\em et~al.\/}(2004)Blijdenstein, van~der Linden, van
  Vliet \& van Aken]{blijdenstein2004scaling}
{\sc \au{Blijdenstein, Theo~BJ}, \au{van~der Linden, Erik}, \au{van Vliet, Ton}
  \& \au{van Aken, George~A}} \yr{2004}  \at{Scaling behavior of delayed
  demixing, rheology, and microstructure of emulsions flocculated by depletion
  and bridging}.  \jt{Langmuir}  \bvol{20}~(26),  \pg{11321--11328}.

\bibitem[Boniello {\em et~al.\/}(2015)Boniello, Blanc, Fedorenko, Medfai,
  Mbarek, In, Gross, Stocco \& Nobili]{boniello2015brownian}
{\sc \au{Boniello, Giuseppe}, \au{Blanc, Christophe}, \au{Fedorenko, Denys},
  \au{Medfai, Mayssa}, \au{Mbarek, Nadia~Ben}, \au{In, Martin}, \au{Gross,
  Michel}, \au{Stocco, Antonio} \& \au{Nobili, Maurizio}} \yr{2015}
  \at{Brownian diffusion of a partially wetted colloid}.  \jt{Nature materials}
   \bvol{14}~(9),  \pg{908}.

\bibitem[Brambilla {\em et~al.\/}(2011)Brambilla, Buzzaccaro, Piazza, Berthier
  \& Cipelletti]{brambilla2011highly}
{\sc \au{Brambilla, Giovanni}, \au{Buzzaccaro, Stefano}, \au{Piazza, R},
  \au{Berthier, Ludovic} \& \au{Cipelletti, Luca}} \yr{2011}  \at{Highly
  nonlinear dynamics in a slowly sedimenting colloidal gel}.  \jt{Physical
  review letters}  \bvol{106}~(11),  \pg{118302}.

\bibitem[Buscall \& White(1987)]{buscall1987consolidation}
{\sc \au{Buscall, Richard} \& \au{White, Lee~R}} \yr{1987}  \at{The
  consolidation of concentrated suspensions. part 1.â€”the theory of
  sedimentation}.  \jt{Journal of the Chemical Society, Faraday Transactions 1:
  Physical Chemistry in Condensed Phases}  \bvol{83}~(3),  \pg{873--891}.

\bibitem[Clark \& Carper(1987)]{growthclark1987phase}
{\sc \au{Clark, John~I} \& \au{Carper, Deborah}} \yr{1987}  \at{Phase
  separation in lens cytoplasm is genetically linked to cataract formation in
  the philly mouse}.  \jt{Proceedings of the National Academy of Sciences}
  \bvol{84}~(1),  \pg{122--125}.

\bibitem[Falconer(2004)]{falconer2004fractal}
{\sc \au{Falconer, Kenneth}} \yr{2004} {\em Fractal geometry: mathematical
  foundations and applications\/}.  \publ{John Wiley \& Sons}.

\bibitem[Fiore {\em et~al.\/}(2017)Fiore, Balboa~Usabiaga, Donev \&
  Swan]{fiore2017rapid}
{\sc \au{Fiore, Andrew~M}, \au{Balboa~Usabiaga, Florencio}, \au{Donev,
  Aleksandar} \& \au{Swan, James~W}} \yr{2017}  \at{Rapid sampling of
  stochastic displacements in brownian dynamics simulations}.  \jt{The Journal
  of Chemical Physics}  \bvol{146}~(12),  \pg{124116}.

\bibitem[Fiore \& Swan(2018{\natexlab{{\em a\/}}})]{fiore2018stokes}
{\sc \au{Fiore, Andrew~M} \& \au{Swan, James~W}} \yr{2018{\natexlab{{\em
  a\/}}}}  \at{Fast stokesian dynamics}.  \jt{In Preparation} .

\bibitem[Fiore \& Swan(2018{\natexlab{{\em b\/}}})]{fiore2018rapid}
{\sc \au{Fiore, Andrew~M} \& \au{Swan, James~W}} \yr{2018{\natexlab{{\em
  b\/}}}}  \at{Rapid sampling of stochastic displacements in brownian dynamics
  simulations with stresslet constraints}.  \jt{The Journal of chemical
  physics}  \bvol{148}~(4),  \pg{044114}.

\bibitem[Gaponik {\em et~al.\/}(2011)Gaponik, Herrmann \&
  EychmuÌˆller]{Gaponik2011}
{\sc \au{Gaponik, Nikolai}, \au{Herrmann, Anne-Kristin} \& \au{EychmuÌˆller,
  Alexander}} \yr{2011}  \at{Colloidal nanocrystal-based gels and aerogels:
  material aspects and application perspectives}.  \jt{The Journal of Physical
  Chemistry Letters}  \bvol{3}~(1),  \pg{8--17}.

\bibitem[Goddard \& Acrivos(1966)]{goddard1966asymptotic}
{\sc \au{Goddard, JD} \& \au{Acrivos, Andreas}} \yr{1966}  \at{Asymptotic
  expansions for laminar forced-convection heat and mass transfer part 2.
  boundary-layer flows}.  \jt{Journal of Fluid Mechanics}  \bvol{24}~(2),
  \pg{339--366}.

\bibitem[Gopalakrishnan {\em et~al.\/}(2006)Gopalakrishnan, Schweizer \&
  Zukoski]{gopalakrishnan2006linking}
{\sc \au{Gopalakrishnan, V}, \au{Schweizer, KS} \& \au{Zukoski, CF}} \yr{2006}
  \at{Linking single particle rearrangements to delayed collapse times in
  transient depletion gels}.  \jt{Journal of Physics: Condensed Matter}
  \bvol{18}~(50),  \pg{11531}.

\bibitem[Harich {\em et~al.\/}(2016)Harich, Blythe, Hermes, Zaccarelli,
  Sederman, Gladden \& Poon]{harich2016gravitational}
{\sc \au{Harich, Rim}, \au{Blythe, TW}, \au{Hermes, Michiel}, \au{Zaccarelli,
  Emanuela}, \au{Sederman, AJ}, \au{Gladden, Lynn~F} \& \au{Poon, WCK}}
  \yr{2016}  \at{Gravitational collapse of depletion-induced colloidal gels}.
  \jt{Soft matter}  \bvol{12}~(19),  \pg{4300--4308}.

\bibitem[Heyes \& Melrose(1993)]{Heyes1993}
{\sc \au{Heyes, DM} \& \au{Melrose, JR}} \yr{1993}  \at{Brownian dynamics
  simulations of model hard-sphere suspensions}.  \jt{Journal of non-newtonian
  fluid mechanics}  \bvol{46}~(1),  \pg{1--28}.

\bibitem[Hu {\em et~al.\/}(2012)Hu, Liao, Chen, Cai, Meng, Liu, Lv, Liu \&
  Yuan]{hu2012novel}
{\sc \au{Hu, Zhenhua}, \au{Liao, Meiling}, \au{Chen, Yinghui}, \au{Cai,
  Yunpeng}, \au{Meng, Lele}, \au{Liu, Yajun}, \au{Lv, Nan}, \au{Liu, Zhenguo}
  \& \au{Yuan, Weien}} \yr{2012}  \at{A novel preparation method for silicone
  oil nanoemulsions and its application for coating hair with silicone}.
  \jt{International journal of nanomedicine}  \bvol{7},  \pg{5719}.

\bibitem[Huh {\em et~al.\/}(2007)Huh, Lynch \& Furst]{huh2007microscopic}
{\sc \au{Huh, Ji~Yeon}, \au{Lynch, Matthew~L} \& \au{Furst, Eric~M}} \yr{2007}
  \at{Microscopic structure and collapse of depletion-induced gels in
  vesicle-polymer mixtures}.  \jt{Physical Review E}  \bvol{76}~(5),
  \pg{051409}.

\bibitem[Ide \& Sornette(2002)]{ide2002oscillatory}
{\sc \au{Ide, Kayo} \& \au{Sornette, Didier}} \yr{2002}  \at{Oscillatory
  finite-time singularities in finance, population and rupture}.  \jt{Physica
  A: Statistical Mechanics and its Applications}  \bvol{307}~(1),
  \pg{63--106}.

\bibitem[Kamp \& Kilfoil(2009)]{kamp2009universal}
{\sc \au{Kamp, Stephen~W} \& \au{Kilfoil, Maria~L}} \yr{2009}  \at{Universal
  behaviour in the mechanical properties of weakly aggregated colloidal
  particles}.  \jt{Soft Matter}  \bvol{5}~(12),  \pg{2438--2447}.

\bibitem[Kilfoil {\em et~al.\/}(2003)Kilfoil, Pashkovski, Masters \&
  Weitz]{kilfoil2003dynamics}
{\sc \au{Kilfoil, Maria~L}, \au{Pashkovski, Eugene~E}, \au{Masters, James~A} \&
  \au{Weitz, DA}} \yr{2003}  \at{Dynamics of weakly aggregated colloidal
  particles}.  \jt{Philosophical Transactions of the Royal Society of London A:
  Mathematical, Physical and Engineering Sciences}  \bvol{361}~(1805),
  \pg{753--766}.

\bibitem[Kim {\em et~al.\/}(2013)Kim, Fang, Eberle, Casta{\~n}eda-Priego \&
  Wagner]{kim2013gel}
{\sc \au{Kim, Jung~Min}, \au{Fang, Jun}, \au{Eberle, Aaron~PR},
  \au{Casta{\~n}eda-Priego, Ram{\'o}n} \& \au{Wagner, Norman~J}} \yr{2013}
  \at{Gel transition in adhesive hard-sphere colloidal dispersions: The role of
  gravitational effects}.  \jt{Physical review letters}  \bvol{110}~(20),
  \pg{208302}.

\bibitem[Kramers(1940)]{kramers1940brownian}
{\sc \au{Kramers, Hendrik~Anthony}} \yr{1940}  \at{Brownian motion in a field
  of force and the diffusion model of chemical reactions}.  \jt{Physica}
  \bvol{7}~(4),  \pg{284--304}.

\bibitem[Lu {\em et~al.\/}(2006)Lu, Conrad, Wyss, Schofield \& Weitz]{Lu2006}
{\sc \au{Lu, Peter~J}, \au{Conrad, Jacinta~C}, \au{Wyss, Hans~M},
  \au{Schofield, Andrew~B} \& \au{Weitz, David~A}} \yr{2006}  \at{Fluids of
  clusters in attractive colloids}.  \jt{Physical review letters}
  \bvol{96}~(2),  \pg{028306}.

\bibitem[Lu {\em et~al.\/}(2008)Lu, Zaccarelli, Ciulla, Schofield, Sciortino \&
  Weitz]{lu2008gelation}
{\sc \au{Lu, Peter~J}, \au{Zaccarelli, Emanuela}, \au{Ciulla, Fabio},
  \au{Schofield, Andrew~B}, \au{Sciortino, Francesco} \& \au{Weitz, David~A}}
  \yr{2008}  \at{Gelation of particles with short-range attraction}.
  \jt{Nature}  \bvol{453}~(7194),  \pg{499}.

\bibitem[MacMinn {\em et~al.\/}(2016)MacMinn, Dufresne \&
  Wettlaufer]{macminn2016large}
{\sc \au{MacMinn, Christopher~W}, \au{Dufresne, Eric~R} \& \au{Wettlaufer,
  John~S}} \yr{2016}  \at{Large deformations of a soft porous material}.
  \jt{Physical Review Applied}  \bvol{5}~(4),  \pg{044020}.

\bibitem[Manley {\em et~al.\/}(2005)Manley, Skotheim, Mahadevan \&
  Weitz]{manley2005gravitational}
{\sc \au{Manley, Suliana}, \au{Skotheim, JM}, \au{Mahadevan, L} \& \au{Weitz,
  DAVID~A}} \yr{2005}  \at{Gravitational collapse of colloidal gels}.
  \jt{Physical review letters}  \bvol{94}~(21),  \pg{218302}.

\bibitem[Mezzenga {\em et~al.\/}(2005)Mezzenga, Schurtenberger, Burbidge \&
  Martin]{mezzenga2005understanding}
{\sc \au{Mezzenga, Raffaele}, \au{Schurtenberger, Peter}, \au{Burbidge, Adam}
  \& \au{Martin, Michel}} \yr{2005}  \at{Understanding foods as soft
  materials}.  \jt{Nature materials}  \bvol{4}~(10),  \pg{729}.

\bibitem[Noro \& Frenkel(2000)]{noro2000extended}
{\sc \au{Noro, Massimo~G} \& \au{Frenkel, Daan}} \yr{2000}  \at{Extended
  corresponding-states behavior for particles with variable range attractions}.
   \jt{The Journal of Chemical Physics}  \bvol{113}~(8),  \pg{2941--2944}.

\bibitem[Northcott {\em et~al.\/}(2005{\natexlab{{\em a\/}}})Northcott, Snape,
  Scales \& Stevens]{northcott2005contaminated}
{\sc \au{Northcott, Kathy~A}, \au{Snape, Ian}, \au{Scales, Peter~J} \&
  \au{Stevens, Geoff~W}} \yr{2005{\natexlab{{\em a\/}}}}  \at{Contaminated
  water treatment in cold regions: an example of coagulation and dewatering
  modelling in antarctica}.  \jt{Cold regions science and technology}
  \bvol{41}~(1),  \pg{61--72}.

\bibitem[Northcott {\em et~al.\/}(2005{\natexlab{{\em b\/}}})Northcott, Snape,
  Scales \& Stevens]{northcott2005dewatering}
{\sc \au{Northcott, Kathy~A}, \au{Snape, Ian}, \au{Scales, Peter~J} \&
  \au{Stevens, Geoff~W}} \yr{2005{\natexlab{{\em b\/}}}}  \at{Dewatering
  behaviour of water treatment sludges associated with contaminated site
  remediation in antarctica}.  \jt{Chemical Engineering Science}
  \bvol{60}~(24),  \pg{6835--6843}.

\bibitem[Padmanabhan \& Zia(2018)]{padmanabhan2018gravitational}
{\sc \au{Padmanabhan, Poornima} \& \au{Zia, Roseanna}} \yr{2018}
  \at{Gravitational collapse of colloidal gels: Non-equiliibrium phase
  separation driven by osmotic pressure}.  \jt{Soft Matter} .

\bibitem[Poon(2002)]{poon2002physics}
{\sc \au{Poon, WCK}} \yr{2002}  \at{The physics of a model colloid--polymer
  mixture}.  \jt{Journal of Physics: Condensed Matter}  \bvol{14}~(33),
  \pg{R859}.

\bibitem[Poon {\em et~al.\/}(1997)Poon, Pirie, Haw \& Pusey]{Poon1997}
{\sc \au{Poon, WCK}, \au{Pirie, AD}, \au{Haw, MD} \& \au{Pusey, PN}} \yr{1997}
  \at{Non-equilibrium behaviour of colloid-polymer mixtures}.  \jt{Physica A:
  Statistical Mechanics and its Applications}  \bvol{235}~(1),  \pg{110--119}.

\bibitem[Poon {\em et~al.\/}(1999)Poon, Starrs, Meeker, Moussaid, Evans, Pusey
  \& Robins]{poon1999delayed}
{\sc \au{Poon, WCK}, \au{Starrs, L}, \au{Meeker, SP}, \au{Moussaid, A},
  \au{Evans, RML}, \au{Pusey, PN} \& \au{Robins, MM}} \yr{1999}  \at{Delayed
  sedimentation of transient gels in colloid--polymer mixtures: dark-field
  observation, rheology and dynamic light scattering studies}.  \jt{Faraday
  Discussions}  \bvol{112},  \pg{143--154}.

\bibitem[Razali {\em et~al.\/}(2017)Razali, Fullerton, Turci, Hallett, Jack \&
  Royall]{razali2017effects}
{\sc \au{Razali, Azaima}, \au{Fullerton, Christopher~J}, \au{Turci, Francesco},
  \au{Hallett, James~E}, \au{Jack, Robert~L} \& \au{Royall, C~Patrick}}
  \yr{2017}  \at{Effects of vertical confinement on gelation and sedimentation
  of colloids}.  \jt{Soft Matter}  \bvol{13}~(17),  \pg{3230--3239}.

\bibitem[Rotne \& Prager(1969)]{rotne1969variational}
{\sc \au{Rotne, Jens} \& \au{Prager, Stephen}} \yr{1969}  \at{Variational
  treatment of hydrodynamic interaction in polymers}.  \jt{The Journal of
  Chemical Physics}  \bvol{50}~(11),  \pg{4831--4837}.

\bibitem[Russel {\em et~al.\/}(1989)Russel, Saville \&
  Schowalter]{russel1989colloidal}
{\sc \au{Russel, William~Bailey}, \au{Saville, Dudley~Albert} \&
  \au{Schowalter, William~Raymond}} \yr{1989} {\em Colloidal dispersions\/}.
  \publ{Cambridge university press}.

\bibitem[Secchi {\em et~al.\/}(2014)Secchi, Buzzaccaro \&
  Piazza]{secchi2014time}
{\sc \au{Secchi, Eleonora}, \au{Buzzaccaro, Stefano} \& \au{Piazza, Roberto}}
  \yr{2014}  \at{Time-evolution scenarios for short-range depletion gels
  subjected to the gravitational stress}.  \jt{Soft Matter}  \bvol{10}~(29),
  \pg{5296--5310}.

\bibitem[Senis {\em et~al.\/}(2001)Senis, Talini \& Allain]{senis2001settling}
{\sc \au{Senis, D}, \au{Talini, L} \& \au{Allain, C}} \yr{2001}  \at{Settling
  in aggregating colloidal suspensions}.  \jt{Oil \& Gas Science and
  Technology}  \bvol{56}~(2),  \pg{153--159}.

\bibitem[Starrs {\em et~al.\/}(2002)Starrs, Poon, Hibberd \&
  Robins]{starrs2002collapse}
{\sc \au{Starrs, LAURA}, \au{Poon, WCK}, \au{Hibberd, DJ} \& \au{Robins, MM}}
  \yr{2002}  \at{Collapse of transient gels in colloid-polymer mixtures}.
  \jt{Journal of Physics: Condensed Matter}  \bvol{14}~(10),  \pg{2485}.

\bibitem[Swan \& Wang(2016)]{swan2016rapid}
{\sc \au{Swan, James~W} \& \au{Wang, Gang}} \yr{2016}  \at{Rapid calculation of
  hydrodynamic and transport properties in concentrated solutions of colloidal
  particles and macromolecules}.  \jt{Physics of Fluids}  \bvol{28}~(1),
  \pg{011902}.

\bibitem[Teece {\em et~al.\/}(2011)Teece, Faers \& Bartlett]{teece2011ageing}
{\sc \au{Teece, Lisa~J}, \au{Faers, Malcolm~A} \& \au{Bartlett, Paul}}
  \yr{2011}  \at{Ageing and collapse in gels with long-range attractions}.
  \jt{Soft Matter}  \bvol{7}~(4),  \pg{1341--1351}.

\bibitem[Teece {\em et~al.\/}(2014)Teece, Hart, Hsu, Gilligan, Faers \&
  Bartlett]{teece2014gels}
{\sc \au{Teece, Lisa~J}, \au{Hart, James~M}, \au{Hsu, Kerry Yen~Ni},
  \au{Gilligan, Stephen}, \au{Faers, Malcolm~A} \& \au{Bartlett, Paul}}
  \yr{2014}  \at{Gels under stress: The origins of delayed collapse}.
  \jt{Colloids and Surfaces A: Physicochemical and Engineering Aspects}
  \bvol{458},  \pg{126--133}.

\bibitem[Varga \& Swan(2016)]{varga2016hydrodynamic}
{\sc \au{Varga, Zsigmond} \& \au{Swan, James}} \yr{2016}  \at{Hydrodynamic
  interactions enhance gelation in dispersions of colloids with short-ranged
  attraction and long-ranged repulsion}.  \jt{Soft matter}  \bvol{12}~(36),
  \pg{7670--7681}.

\bibitem[Varga \& Swan(2018{\natexlab{{\em a\/}}})]{varga2018large}
{\sc \au{Varga, Zsigmond} \& \au{Swan, James~W}} \yr{2018{\natexlab{{\em
  a\/}}}}  \at{Large scale anisotropies in sheared colloidal gels}.
  \jt{Journal of Rheology}  \bvol{62}~(2),  \pg{405--418}.

\bibitem[Varga \& Swan(2018{\natexlab{{\em b\/}}})]{varga2017normal}
{\sc \au{Varga, Zsigmond} \& \au{Swan, James~W}} \yr{2018{\natexlab{{\em
  b\/}}}}  \at{Normal modes of weak colloidal gels}.  \jt{Physical Review E}
  \bvol{97}~(1),  \pg{012608}.

\bibitem[Varga {\em et~al.\/}(2015)Varga, Wang \& Swan]{varga2015hydrodynamics}
{\sc \au{Varga, Zsigmond}, \au{Wang, Gang} \& \au{Swan, James}} \yr{2015}
  \at{The hydrodynamics of colloidal gelation}.  \jt{Soft Matter}
  \bvol{11}~(46),  \pg{9009--9019}.

\bibitem[Weeks {\em et~al.\/}(2000)Weeks, van Duijneveldt \&
  Vincent]{weeks2000formation}
{\sc \au{Weeks, James~R}, \au{van Duijneveldt, Jeroen~S} \& \au{Vincent,
  Brian}} \yr{2000}  \at{Formation and collapse of gels of sterically
  stabilized colloidal particles}.  \jt{Journal of Physics: Condensed Matter}
  \bvol{12}~(46),  \pg{9599}.

\bibitem[Yang {\em et~al.\/}(2004)Yang, Zhang, Yang, Chen, Zhuang, Xu \&
  Wang]{yang2004molecularly}
{\sc \au{Yang, Huang-Hao}, \au{Zhang, Shu-Qiong}, \au{Yang, Wei}, \au{Chen,
  Xiao-Lan}, \au{Zhuang, Zhi-Xia}, \au{Xu, Jin-Gou} \& \au{Wang, Xiao-Ru}}
  \yr{2004}  \at{Molecularly imprinted sol- gel nanotubes membrane for
  biochemical separations}.  \jt{Journal of the American Chemical Society}
  \bvol{126}~(13),  \pg{4054--4055}.

\bibitem[Zaccarelli(2007)]{zaccarelli2007colloidal}
{\sc \au{Zaccarelli, Emanuela}} \yr{2007}  \at{Colloidal gels: equilibrium and
  non-equilibrium routes}.  \jt{Journal of Physics: Condensed Matter}
  \bvol{19}~(32),  \pg{323101}.

\end{thebibliography}


\end{document}